\documentclass[twocolumn,preprintnumbers,amsmath,amssymb,nofootinbib]{revtex4}

\usepackage{amssymb}
\usepackage{bm}
\usepackage[plainpages=false]{hyperref}
\usepackage{graphicx}
\usepackage{multirow}
\usepackage{bbding}

\usepackage{color}
\usepackage[normalem]{ulem}

\begin{document}

\title{Probing Intergalactic Magnetic Fields with Simulations of Electromagnetic Cascades}

\author{Rafael \surname{Alves Batista}$^{\it \bf 1,}$\footnote[1]{E-mail: rafael.alvesbatista@physics.ox.ac.uk}, Andrey Saveliev$^{\it \bf 2,3,}$\footnote[2]{E-mail: andrey.saveliev@desy.de, {\it corresponding author}}, G\"unter Sigl$^{\it \bf 2,}$\footnote[3]{E-mail: guenter.sigl@desy.de}, Tanmay Vachaspati$^{\it \bf 4,}$\footnote[4]{E-mail: tvachasp@asu.edu}}

\affiliation{
$^ {\bf \it 1}$University of Oxford, Department of Physics - Astrophysics, Denys Wilkinson Building, Keble Road, Oxford OX1 3RH, United Kingdom\\
$^ {\bf \it 2}$Universit\"at Hamburg, {II}. Institute for Theoretical Physics, Luruper Chaussee 149, 22761 Hamburg, Germany\\
$^ {\bf \it 3}$Russian Academy of Sciences, Keldysh Institute of Applied Mathematics, Miusskaya sq.~4, 125047 Moscow, Russian Federation\\
$^ {\bf \it 4}$Arizona State University,  Physics Department, 650 E. Tyler Mall, Tempe, AZ 85287, United States of America
}

\begin{abstract}
We determine the effect of intergalactic magnetic fields on the distribution of high energy gamma rays by performing three-dimensional Monte Carlo 
simulations of the development of gamma-ray-induced electromagnetic cascades in the magnetized intergalactic medium. We employ the so-called ``Large Sphere 
Observer'' method to efficiently simulate blazar gamma ray halos. We study magnetic fields with a Batchelor spectrum and with maximal left- 
and right-handed helicities. We also consider the case of sources whose jets are tilted with respect to the line of sight. We verify the formation of extended gamma ray halos around the source direction, and observe spiral-like patterns if the magnetic field is helical. 
We apply the $Q$-statistics to the simulated halos to 
extract their spiral nature and also propose an alternative method, the $S$-statistics. Both methods provide a quantative way to infer the helicity of the intervening 
magnetic fields from the morphology of individual blazar halos for magnetic field strengths $B \gtrsim 10^{-15}\,{\rm G}$ and magnetic coherence lengths 
$L_{\rm c} \gtrsim 100\,{\rm Mpc}$. We show that the $S$-statistics has a better performance than the $Q$-statistics when assessing magnetic helicity from the simulated halos.
\end{abstract}

\maketitle

\section{Introduction}

The origin, strength and structure of intergalactic magnetic fields (IGMF) remain a mystery up to the present day. Possible mechanisms to explain cosmic 
magnetogenesis may be divided into two main categories: cosmological scenarios predict that magnetic fields were generated through processes taking place in 
the early universe, such as inflation \cite{PhysRevD.37.2743,Ratra:1991bn,Byrnes:2011aa,Ferreira:2014hma}, electroweak
\cite{Vachaspati:1991nm,Enqvist:1993np,PhysRevD.53.662,Grasso:1997nx} or QCD phase transitions
\cite{Hogan:1983zz,1989ApJ...344L..49Q,PhysRevD.55.4582,Tevzadze:2012kk}, and leptogenesis \cite{Long:2013tha}, among others; in astrophysical scenarios the 
fields would be created during the later stages of evolution of the universe, for example during structure formation~\cite{Kulsrud:1996km} or even thereafter
\cite{1981ApJ...248...13L}.

Measurements of IGMF are rather difficult due to their low magnitude. Common methods to estimate the strength of IGMF are indirect and include the well-known 
Faraday rotation measurements which yield upper limits of the order of a few nG~\cite{PhysRevD.80.123012}. Lower bounds, $B  \gtrsim 10^{-17}\, \text{G}$, have 
been obtained by several authors using gamma-ray-induced electromagnetic cascades in the intergalactic space \cite{JETPLett.85.10.473,PhysRevD.80.123012,Essey:2010nd,2010ApJ...722L..39A,2010MNRAS.406L..70T,Taylor:2011bn,Takahashi:2013lba,2014arXiv1410.7717C}.
These lower bounds are controversial because of the claims~\cite{0004-637X-752-1-22,0004-637X-758-2-102,Schlickeiser:2013eca,SavelievIGM,Chang:2014cta} that 
the development of the cascade is suppressed by plasma instabilities that arise from interactions with the intergalactic medium. On the other hand, recent 
direct observations of cascades~\cite{Chen11072015} suggest that plasma instabilities are not operative and that the original bounds hold. We expect that 
future analyses will clarify the role, if any, of plasma instabilities in the development of the electromagnetic cascade.

Magnetic fields can carry helicity ($\mathcal{H}$), which is defined as
\begin{equation}
\mathcal{H} = \int \mathbf{A} \cdot \mathbf{B} \,  {\rm d}^{3}r \,,
\end{equation}
where $\mathbf{A}$ is the magnetic vector potential and $\mathbf{B}=\nabla \times \mathbf{A}$ is the magnetic field. Since magnetic helicity affects the 
dynamical evolution of magnetic fields, an indirect way to measure magnetic helicity is to measure the magnetic field power spectrum and compare it with the 
evolution seen in magnetohydrodynamical (MHD) simulations \cite{Sigl:2002kt,0004-637X-640-1-335,Saveliev:2013uva}. Ther are also some proposals to 
{\em directly} measure magnetic helicity based on the propagation of cosmic rays~\cite{Kahniashvili:2005yp}. More recently, it has been proposed that 
helicity can leave characteristic parity-odd imprints on the arrival directions of gamma rays that are the result of gamma-ray-induced electromagnetic cascades
\cite{PhysRevD.87.123527,Tashiro:2013ita,Tashiro:2014gfa,Long:2015bda,Chen11072015}. In particular, Long \& Vachaspati~\cite{Long:2015bda} have carried out a 
thorough analysis of the morphology of the arrival directions of gamma rays using a semi-analytical approach, but without including the stochasticity of the 
magnetic field or the cascade process. Hence, a full Monte Carlo approach and three-dimensional simulations are needed in order to confirm or refute their 
findings and provide a solid basis for further analyses. 

The observation of helical primordial magnetic fields has profound implications for particle physics and the early universe. Scenarios in which the 
cosmological matter-antimatter asymmetry is generated dynamically are found to also produce helical magnetic fields~\cite{Vachaspati:2001nb}. The handedness of 
the field is related to details of the matter-genesis scenario~\cite{Vachaspati:2001nb,Long:2013tha}. If the observed magnetic fields are coherent on very 
large scales, they may have been produced at the initial epoch, perhaps during an inflation \cite{PhysRevD.37.2743,Ratra:1991bn}. Helicity on these 
scales would indicate the presence of certain parity violating interactions in the fundamental Lagrangian~\cite{Caprini:2014mja}.

In the present work we perform simulations of the propagation of gamma rays in both helical and non-helical IGMF. This paper is structured as follows: first, 
we discuss the theory and implementation of simulations of electromagnetic cascades in Sec.~\ref{sec:EMSim}; in Sec.~\ref{sec:Results} we apply our approach to 
different magnetic field configurations, focusing in particular on the role of magnetic helicity (Sec.~\ref{sec:HelMag1} - \ref{sec:HelMag3}); in 
Sec.~\ref{sec:Discussion} we discuss the results, draw our conclusions and give a short outlook.

\section{Simulations of Electromagnetic Cascades in the Intergalactic Medium} \label{sec:EMSim}

\subsection{Interactions and Energy Losses}

The basic physics underlying the development of electromagnetic cascades induced by high energy gamma rays from blazars is well-known 
\cite{1994ApJ...423L...5A,Plaga1994}. A gamma ray emitted by a blazar interacts with photons from the diffuse extragalactic background radiation fields 
producing an electron-positron pair. The electrons\footnote{Hereafter we will collectively refer to electrons and positrons simply as ``electrons''.} then 
upscatter photons of the cosmic microwave background to high energies in a process known as inverse Compton scattering (ICS). The electrons continue to 
upscatter photons until their energy diminishes. The upscattered photons can produce yet more electron-positron pairs until the energy of the photon drops 
below the threshold for pair production. We should therefore observe the blazar source as well as gamma rays originating from the cascade process, unless 
magnetic fields bend the electron trajectories sufficiently away from the line of sight.

To perform three-dimensional simulations of the development of gamma-ray-induced electromagnetic cascades in the IGM, we have modified the CRPropa 3
\cite{CRPropa} code, commonly used for ultra-high energy cosmic ray propagation. Taking advantage of the modular structure of the code 
and the flexibility to handle custom magnetic field configurations, we have implemented relevant interactions for gamma rays and electrons in the energy range 
of interest ($1\,\mathrm{GeV} \lesssim E \lesssim 1\,\mathrm{PeV}$). Relevant interactions are pair production by gamma rays and inverse Compton scattering by electrons. Adiabatic losses due to the expansion of the universe are also taken into account.  Synchrotron losses, albeit small in this energy range, are considered as well, for the sake of completeness.

Particles are propagated step-by-step. Within each step the probability of a given interaction to occur is computed using  tabulated values for the interaction rate. If the particle is charged, deflections due to magnetic fields are calculated by integrating the equations of motion. By doing so, we are adopting a three-dimensional Monte Carlo approach for the propagation.

Interaction rates for pair production and inverse Compton scattering are calculated following the implementation used in the Elmag code~\cite{2012CoPhC.183.1036K}, and defined as the inverse of the mean free path $\lambda$. They are tabulated for the CMB and various models of extragalactic background light (EBL) at different redshifts as follows\footnote{Unless otherwise stated, in this section we use ``natural units'' in which $\hbar = c = 1$.}~\cite{2012CoPhC.183.1036K}:
\begin{equation}
R(E,z) \equiv \lambda^{-1}(E,z) = \frac{1}{8E^2} \int\limits^\infty_0 {\rm d}\varepsilon \int\limits^{s_{\rm max}}_{s_{\rm min}} {\rm d}s \frac{n(\varepsilon, z)}{\varepsilon^2} F_{\rm int} (s)  \,, 
\end{equation}
where $E$ is the energy of the interacting particle (electron, positron or photon), $n(\varepsilon, z)$ is the comoving spectral density distribution of photons with energy $\varepsilon$ at redshift $z$, $s$ denotes the center of mass energy in the kinematic range $s_{\rm min} \leq s \leq s_{\rm max}$, and $F_{\rm int}$ is a function that depends on the interaction in question. 

In the case of pair production $F_{\rm int} = F_{\rm PP}$ is
\begin{equation}
F_{\rm PP}(s) = s \sigma_{\rm PP}(s)\,, 
\end{equation}
where $\sigma_{\rm PP}(s)$ is the cross section for pair production, and $s = 2 E \varepsilon (1 - \cos\theta)$, with $0 \leq \theta \leq \pi$ being the angle between the gamma ray of energy $E$ and the background photon of energy $\varepsilon$. The values of $s$ range from $s_{\rm min} = 4 m_e^2$ to $s_{\rm max} = 4 E \varepsilon_{\rm max}$, where $\varepsilon_{\rm max}$ is the cutoff energy for the photon field, assumed to be approximately 0.1 eV for the CMB and 15 eV for the EBL. 

For inverse Compton scattering $F_{\rm int} = F_{\rm ICS}$ is given by
\begin{equation}
F_{\rm ICS}(s) = \frac{1}{\beta} \sigma_{\rm ICS}(s - m_e^2)\,,
\end{equation}
with $\beta = ( 1 - m_{e}^2 / E^{2})^{\frac{1}{2}}$. The center of mass energies in this case are $s = m_e^2 + 2 E \epsilon (1 - \beta \cos\theta)$, for $s_{\rm min} \leq s \leq s_{\rm max}$, with $s_{\rm min} = m_{e}^{2}$ and $s_{\rm max} = m_{e}^{2} + 2 E \epsilon_{\rm max} (1 + \beta)$. 

Cross sections for these interactions are well-known (see e.g.~\cite{jauch1976,2012CoPhC.183.1036K}). The spectral density distribution of the cosmic microwave background (CMB) can be described as a black-body. The EBL is model-dependent. For this particular work we adopt the lower limit EBL model of Kneiske \& Dole~\cite{2010A&A...515A..19K}.

Synchrotron losses are given by
\begin{equation}
\frac{{\rm d}E}{{\rm d}x} = \frac{m_e^2 \chi^2}{(1 + 4.8(1 + \chi) \ln(1 + 1.7 \chi) + 3.4 \chi^2)^{2/3}}\,,
\end{equation}
following Ref.~\cite{2012CoPhC.183.1036K}. Here $m_{e}$ is the electron mass, $\chi$ is
\begin{equation}
\chi \equiv \frac{\left| \mathbf{p} \times \mathbf{B} \right| } {m_{e} B_{0}}\,,
\end{equation}
with $B_{0} = 4.1 \times 10^{13}\,$G, and $\mathbf{B}$ the magnetic field vector acting on an electron with momentum $\mathbf{p}$.

Adiabatic losses due to the expansion of the universe are given by
\begin{equation}
-\frac{1}{E}\frac{{\rm d}E}{{\rm d}x} =  \frac{H(t)}{c} = \frac{H_0}{c} \sqrt{\Omega_{\rm m} (1+z)^3 +  \Omega_{\Lambda}}\,,
\end{equation}
with $H_0 \equiv H(0) \simeq 70\,$km/s/Mpc designating the Hubble constant at present time, $\Omega_{\rm m} \simeq 0.3$ being the density of matter, and $\Omega_{\Lambda} \simeq 0.7$ being the density of dark energy, assuming the standard $\Lambda$CDM cosmological model.

In our simulations we consider a monochromatic source and all emitted gamma rays are assumed to have an energy of $10\,{\rm TeV}$. Photons from the source with energies much smaller than this will be below the threshold for creating a cascade, while photons with much 
higher energies will have a diminished flux.

\subsection{Sampling of Helical Magnetic Fields} 

In order to run a simulation for a given magnetic field scenario or, more specifically, for a given magnetic field (and magnetic helicity) spectrum, one has to sample a magnetic field grid which then may be used as input. This procedure is explained in the following using the formalism of~\cite{Tashiro:2012mf}.

The aim is to decompose the magnetic field into modes of the divergence-free eigenfunctions $\mathbf{K}^{\pm}$ of the Laplace operator which for a specific wave vector $\mathbf{k}$ are given by\footnote{We adopt CGS units in this section.}
\begin{equation} \label{Qpm}
\mathbf{K}^{\pm}(\mathbf{k}) = \mathbf{e}^{\pm}(\mathbf{k}) e^{i \mathbf{k} \cdot \mathbf{x}} \equiv \frac{\mathbf{e}_{1}(\mathbf{k}) \pm i \mathbf{e}_{2}(\mathbf{k})}{\sqrt{2}} e^{i \mathbf{k} \cdot \mathbf{x}}\,,
\end{equation}
where $(\mathbf{e}_{1}, \mathbf{e}_{2}, \mathbf{e}_{3})$ is a right-handed orthonormal system of real unit vectors with 
$\mathbf{e}_{3} = \mathbf{k}/k \equiv \hat{\mathbf{k}}$. In order to obtain $\mathbf{e}_{1}$ and $\mathbf{e}_{2}$ we chose a fixed arbitrary vector 
$\mathbf{n}_{0} \nparallel \mathbf{k}$ with which we calculate
\begin{equation}
\mathbf{e}_{1} \equiv \frac{\mathbf{n}_{0} \times \hat{\mathbf{k}}}{\left| \mathbf{n}_{0} \times \hat{\mathbf{k}} \right|}\, , \, \mathbf{e}_{2} \equiv \frac{\hat{\mathbf{k}} \times \mathbf{e}_{1}}{\left| \hat{\mathbf{k}} \times \mathbf{e}_{1} \right|}\,.
\end{equation}
With these definitions the $\mathbf{K}^{\pm}$ fullfil the following relations~\cite{Tashiro:2012mf}:
\begin{equation} \label{nablaQpm}
\nabla \cdot \mathbf{K}^{\pm} = 0\, , \, \nabla \times \mathbf{K}^{\pm} = \pm k \mathbf{K}^{\pm}\,.
\end{equation}

Considering these relations the magnetic field with $\nabla \cdot \mathbf{B} =0$ or, in Fourier space, $\mathbf{k} \cdot \tilde{\mathbf{B}}(\mathbf{k}) = 0$ may be decomposed as
\begin{equation} \label{Bx}
\mathbf{B}(\mathbf{x}) = \int\left[ \tilde{B}^{+}(\mathbf{k}) \mathbf{K}^{+}(\mathbf{k}) + \tilde{B}^{-}(\mathbf{k}) \mathbf{K}^{-}(\mathbf{k}) \right] \frac{{\rm d}^{3}k}{(2 \pi)^{3}} \,,
\end{equation}
for which, in order for $\mathbf{B}(\mathbf{x})$ to be real, the condition
\begin{equation}
\begin{split}
&\tilde{B}^{+}(\mathbf{k}) \mathbf{e}^{+}(\mathbf{k}) + \tilde{B}^{-}(\mathbf{k}) \mathbf{e}^{-}(\mathbf{k}) \\
&= \tilde{B}^{+}(-\mathbf{k})^{*} \mathbf{e}^{+}(-\mathbf{k})^{*} + \tilde{B}^{-}(-\mathbf{k})^{*} \mathbf{e}^{-}(-\mathbf{k})^{*}
\end{split}
\end{equation}
must hold. A possible realization of this condition is
\begin{equation} \label{Breal}
\tilde{B}^{\pm}(\mathbf{k}) \mathbf{e}^{\pm}(\mathbf{k}) = \tilde{B}^{\pm}(-\mathbf{k})^{*} \mathbf{e}^{\pm}(-\mathbf{k})^{*}\,,
\end{equation}
which can be fulfilled by setting
\begin{equation} \label{Bcondition}
\tilde{B}^{\pm}(\mathbf{k}) = \tilde{B}^{\pm}(-\mathbf{k})^{*}\,,
\end{equation}
which we are going to use in the following. Together with (\ref{Qpm}) and (\ref{Breal}) this leads to
\begin{equation}
\mathbf{e}^{\pm}(\mathbf{k}) = \mathbf{e}^{\pm}(-\mathbf{k})^{*},
\end{equation}
and thus
\begin{equation}
\mathbf{e}_{1}(\mathbf{k}) = - \mathbf{e}_{1}(- \mathbf{k})\, , \, \mathbf{e}_{2}(\mathbf{k}) = \mathbf{e}_{2}(- \mathbf{k}) \,.
\end{equation}

The $\tilde{B}^{\pm}$ may be obtained from the given spectra using the relations \cite{Tashiro:2012mf}
\begin{equation} \label{EB(k)}
\begin{aligned}
\frac{1}{8 \pi} \langle \left| \mathbf{B}(\mathbf{x}) \right|^2) \rangle {} & = \int \left[ \left| \tilde{B}^{+} (\mathbf{k}) \right|^2  + \left| \tilde{B}^{-} (\mathbf{k}) \right|^2 \right] \frac{k^{2} \, {\rm d} k}{16 \pi^{3}} \\ 
\\ 
& \equiv \int E_{B}(k) \, {\rm d} \ln k
\end{aligned}
\end{equation}
and
\begin{equation} \label{helicitymodes}
\begin{aligned}
	\langle \mathbf{A}(\mathbf{x}) \cdot \mathbf{B}(\mathbf{x}) \rangle {} 
	& = \int \left[ \left| \tilde{B}^{+} (\mathbf{k}) \right|^2 - \left|\tilde{B}^{-} (\mathbf{k}) \right|^2 \right] \frac{k {\rm d} k}{2 \pi^{2}} \\  & \equiv \int H_{B}(k) \, {\rm d} \ln k\,,
\end{aligned}
\end{equation}
where $\mathbf{A}$ is the vector potential and $E_{B}$ and $H_{B}$ are the spectra of the magnetic energy density and the magnetic helicity density, respectively. $E_{B}$ and $H_{B}$ are related to each other through the inequality~\cite{Brandenburg20051}
\begin{equation}
\frac{k}{8 \pi} \left| H_{B}(k) \right| \le E_{B}(k)\,
\end{equation}
which may be also expressed as
\begin{equation} \label{HBEB}
H_{B}(k) = f_{H}(k) \frac{8 \pi}{k} E_{B}(k)
\end{equation}
with $-1 \le f_{H}(k) \le 1$. 

Numerical and analytical analyses \cite{Kahniashvili:2012uj,Saveliev:2013uva}
show that $E_{B}$ is a power-law for small $k$, {\it i.e.}
\begin{equation} \label{EBkalpha}
E_{B} \propto k^{\alpha}\,,
\end{equation}
with $\alpha = 5$. This power-law behavior for $E_B$ is also known as the Batchelor spectrum.

In our numerical analysis with stochastic magnetic fields of Sec.~\ref{sec:HelMag1}, we will use magnetic fields with the spectrum 
\begin{equation}
E_{B} \propto 
\begin{cases}
k^{5} \,,\, k \leq 2 \pi/ L_{\rm min}\,,\\
0 \,,\, k > 2 \pi/ L_{\rm min}\,,
\end{cases}
\label{EB120}
\end{equation}
where, for a correlation length $L_{\rm c} = 120\,{\rm Mpc}$, $L_{\rm min} = 8L_{\rm c}/5 = 192\,{\rm Mpc}$
is the cutoff scale (cf.~Eq.~(\ref{Lcdef}) below). Finally, solving (\ref{EB(k)}) and (\ref{helicitymodes}) for 
$\left|\tilde{B}^{\pm}\right|^{2}$ gives
\begin{equation} \label{BsquaredEB}
\begin{split}
\left|\tilde{B}^{\pm}\right|^{2} &= \frac{8 \pi^{3}}{k^{3}} \left[ E_{B}(k) \pm \frac{k}{8 \pi}H_{B}(k)\right] \\
&= \left(\frac{2 \pi}{k}\right)^{3} \left[ 1 \pm f_{H}(k) \right] E_{B}(k)\,.
\end{split}
\end{equation} 

With these considerations the procedure for sampling a magnetic field for given spectra $E_{B}$ and $H_{B}$ 
on a grid in $x$-space is the following: first, for each $\mathbf{k}$ in the Fourier-transformed $k$-space a value 
for the norm of $\tilde{B}(k)$ is generated from a normal distribution with mean value $\mu = 0$ and standard deviation 
$\sigma =2(2\pi/k)^3 E_B(k)$ as follows from (\ref{BsquaredEB}) with $f_{\rm H} = \pm 1$. Next, we include a random phase factor
\begin{equation}
\tilde{B}^{\pm}(\mathbf{k}) = \left| \tilde{B}^{\pm}(\mathbf{k}) \right| \left[ \cos\theta^{\pm}(\mathbf{k}) + i \sin\theta^{\pm}(\mathbf{k}) \right],
\end{equation}
where $\theta^{\pm}(\mathbf{k})$ are random phases distributed uniformly on $[0;2\pi)$. Once we have $\tilde{B}^{\pm}({\bf k})$, we use Eq.~(\ref{Bcondition}) to 
find $\tilde{B}^{\pm}(-{\bf k})$. These $\tilde{B}^{\pm}(\mathbf{k})$ can then be plugged into (\ref{Bx}) to obtain 
the value for $\mathbf{B}(\mathbf{x})$ at a given $\mathbf{x}$. 

As sometimes it is more convenient to have $\tilde{\mathbf{B}}(\mathbf{k})$ given in terms of the real and imaginary parts, we use Eq.~(\ref{Qpm}) to write it 
down in the form
\begin{widetext}
\begin{equation}
\begin{split}
\tilde{\mathbf{B}}(\mathbf{k}) &= \tilde{B}^{+}(\mathbf{k}) \mathbf{e}^{+}(\mathbf{k}) + \tilde{B}^{-}(\mathbf{k}) \mathbf{e}^{-}(\mathbf{k}) \\
&=
\frac{1}{\sqrt{2}} \Big\{ \left[ \left( \left| \tilde{B}^{+}(\mathbf{k}) \right| \cos\theta^{+} + \left| \tilde{B}^{-}(\mathbf{k}) \right| \cos\theta^{-}\right) \mathbf{e}_{1} + \left( - \left| \tilde{B}^{+}(\mathbf{k}) \right| \sin\theta^{+} + \left| \tilde{B}^{-}(\mathbf{k}) \right| \sin\theta^{-}\right) \mathbf{e}_{2} \right] \\
&+ i \left[ \left( \left| \tilde{B}^{+}(\mathbf{k}) \right| \sin\theta^{+} + \left| \tilde{B}^{-}(\mathbf{k}) \right| \sin\theta^{-}\right) \mathbf{e}_{1} + \left( \left| \tilde{B}^{+}(\mathbf{k}) \right| \cos\theta^{+} - \left| \tilde{B}^{-}(\mathbf{k}) \right| \cos\theta^{-}\right) \mathbf{e}_{2} \right] \Big\}\,.
\end{split}
\end{equation}
\end{widetext}

\section{Results} \label{sec:Results}

In this section we present the results. Some preliminary considerations regarding the setup of simulations should first be made. 

We use the Large Sphere Observer approach which is a computationally efficient method for studying cosmic and gamma rays from a single source~\cite{PhysRevD.80.023010,2041-8205-719-2-L130}. 
It is defined by the fact that this source is located in the center of a sphere which has a radius equal to 
$D_{\rm s}$, the distance from the source to the observer. Hence, if a particle crosses the sphere from the inside to the outside, it is flagged `detected'. This will henceforth be called a `hit' and it corresponds 
to the particle reaching the observer.

The source can emit gamma rays either within a jet or isotropically. Due to the choice of a large sphere as an observer, all events above a given energy threshold (here we
use $1.5\,{\rm GeV}$) are detected. Moreover, we can easily select a subset of the events and consider an 
arbitrary emission pattern, such as a jet of 
arbitrary half-opening angle $\Psi$, or an emission around an arbitrary direction tilted with respect to the line of sight.

Simple geometrical considerations allow us to correct the arrival directions on the large sphere to mimic Earth's 
field of view. In the sky maps presented in this work, for each hit, the corresponding coordinate system of the 
observer is placed such that its origin is located at the position of the hit while the $z$ axis points towards the 
source, {\it i.e.} in the direction of the center of the sphere. 
In order to determine the directions of the $x$ and $y$ axes, we take a ``global'' reference frame at
a fixed point of the sphere and parallel-transport it along a geodesic to the location of the hit.
Then the spherical angles of the event are measured in the local frame located at the hit point.

While the ``Large Sphere Observer'' method is economical as no photons are wasted, 
one possible concern is that in a realistic set-up most photons would indeed be wasted 
and the actual halo morphology would be sensitive to the absent photons. However,
our results in the test case of a uniform magnetic field correlate well with analytic
simulations \cite{Long:2015bda}, giving us confidence in the method.

The magnetic field (except for the uniform case) is sampled in a grid with $1000^{3}$ cells, where each cell 
has a size of $\sim 10~{\rm Mpc}$. 

\subsection{Comparison with Analytic Estimates} \label{sec:GenMag}

\begin{figure*}[ht]
\center
\includegraphics[scale=0.379]{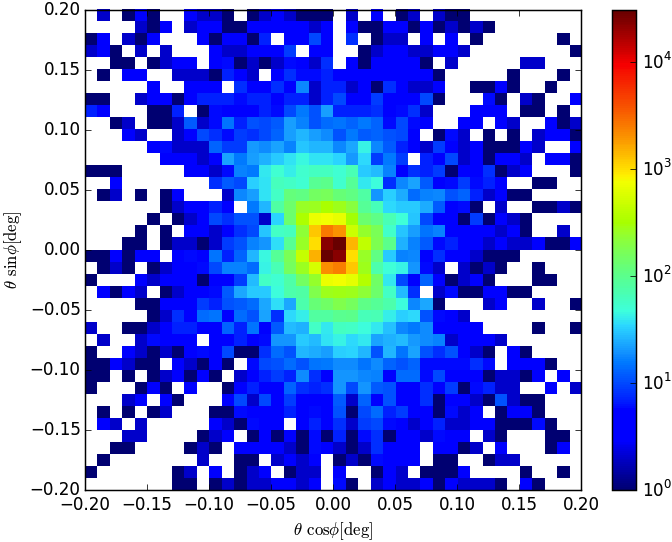}
\includegraphics[scale=0.379]{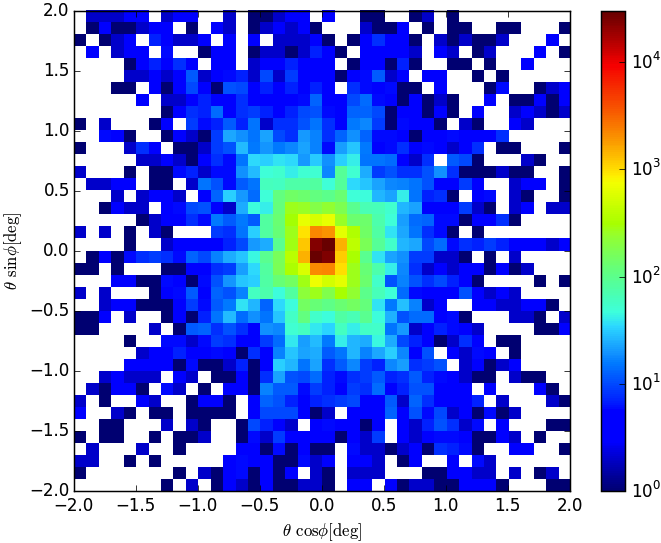}
\includegraphics[scale=0.505]{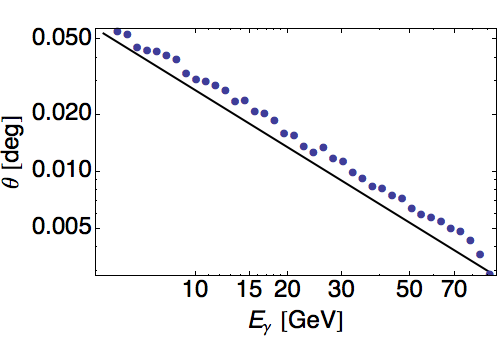}
\includegraphics[scale=0.505]{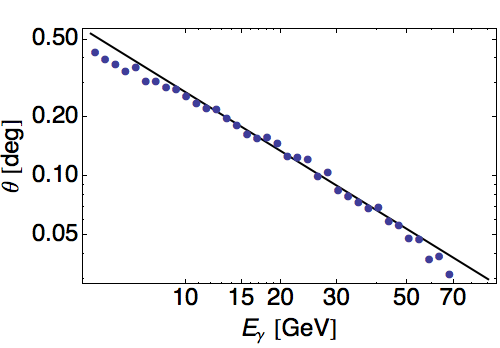}
\caption{Arrival directions of photons from a monochromatic TeV blazar emitting gamma rays with energies $E_{\rm TeV}=10\text{ TeV}$ in a collimated jet, in the energy range of 1-100 GeV, projected onto a plane, are shown in the upper row; the color scale indicates the number of photons per bin. The deflection angle ($\theta$) of observed gamma rays as a function of the energy are presented in the bottom row. The magnetic field is stochastic  with a spectrum according to (\ref{EBkalpha}) and a mean field strength of $10^{-15}\,{\rm G}$ (left column) and $10^{-16}\,{\rm G}$ (right column); blue dots correspond to simulation results and the black line represents the analytical prediction using Eq.~(\ref{thetaEBD}).}
\label{fig:AnalytComp}
\end{figure*}

For a gamma ray emitted at TeV energy $E_{\rm TeV}$ and observed at an energy $E_{\gamma}$, originating 
from a source (blazar in our case) located at redshift $z_{\rm s}$ and distance $D_{\rm s}$, traversing a magnetic 
field of strength $B$, the expected average angular arrival direction is~\cite{PhysRevD.80.123012}
\begin{eqnarray} \label{thetaEBD}
\theta(E_{\gamma}) &\simeq& 0.05^{\circ} \kappa (1 + z_{\rm s})^{-4}  \nonumber \\
&& \hskip -1.5 cm
\times \left( \frac{B}{{\rm fG}} \right) \left( \frac{E_{\gamma}}{0.1\,{\rm TeV}} \right)^{-1} 
\left( \frac{D_{\rm s}}{{\rm Gpc}} \right)^{-1} \left( \frac{E_{\rm TeV}}{10 \,\rm{TeV}} \right)^{-1}.
\end{eqnarray}
This formula is only a rough estimate  where $\kappa$ is a factor close to unity, $\kappa \simeq 1$, which 
varies slightly with the EBL model chosen.
Furthermore, this equation is only valid if the coherence length ($L_{\rm c}$) of the field is much larger than 
the propagation length of electrons before they upscatter photons via inverse Compton. This is always
true in our simulations because the propagation length is of the order of 30~kpc, whereas the minimum coherence length is 10~Mpc.

In order to compare our results with Eq.~(\ref{thetaEBD}), we simulate the propagation of gamma rays with initial energies $E_{\rm TeV} = 10\,{\rm TeV}$, 
distance $D_{\rm s} = 1\,{\rm Gpc}$ ($z_s \simeq 0.25$), emitted in a collimated jet along the line of sight assuming stochastic magnetic fields with strength 
of $B = 10^{-16}\,{\rm G}$ and $B = 10^{-15}\,{\rm G}$. The maps containing the arrival directions are shown in the top panel of Fig.~\ref{fig:AnalytComp}. 

We have compared the deflections obtained from the simulations with the theoretical prediction of Eq.~(\ref{thetaEBD}). This is shown in the bottom panel of Fig.~\ref{fig:AnalytComp}. The results show a good agreement with the expected deflections. Differences are due to the nature of the analytic formula itself, which has been derived in~\cite{PhysRevD.80.123012} using various simplifying assumptions. Furthermore, as has been pointed out in Ref.~\cite{PhysRevD.80.123012}, the deflection angle is highly sensitive to the particular EBL model used. In particular, for the EBL model used here (Kneiske \& Dole~\cite{2010A&A...515A..19K}), we expect $\kappa \approx 2.3$. As pointed out in Ref.~\cite{PhysRevD.80.123012}, $0.3 \lesssim \kappa \lesssim 3.0$ for typical EBL models found in the literature.

\subsection{Uniform Magnetic Fields} \label{sec:HomMag}

We now consider a simple scenario with a uniform magnetic field. By definition, a uniform magnetic field  has a preferred direction, and therefore one has to distinguish among three general cases depending on the orientation of the magnetic field with respect to the axis of the jet: parallel, perpendicular, and intermediate orientation. The jet direction is assumed to be along the line of sight.

\begin{figure*}
\center
\includegraphics[scale=0.268]{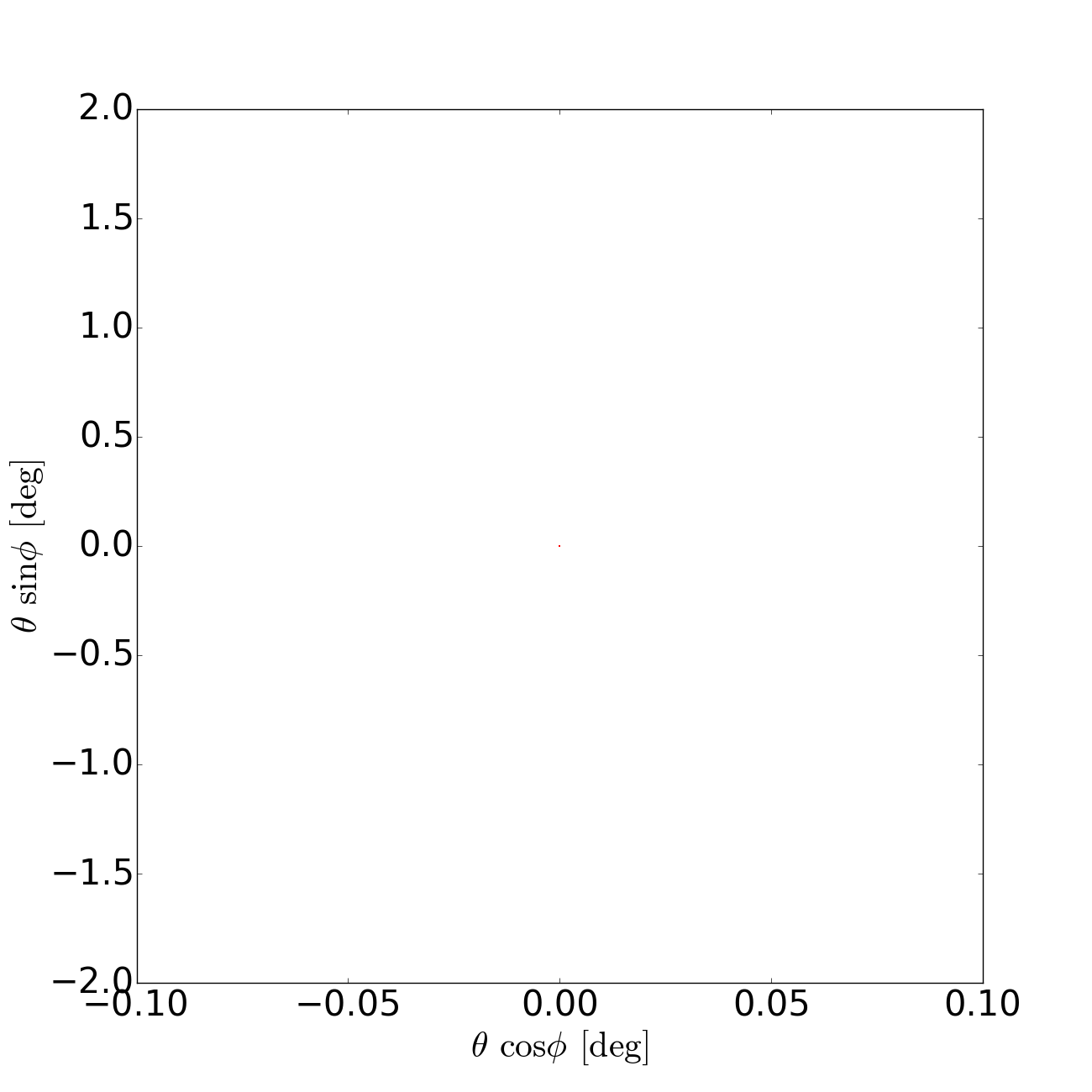}
\includegraphics[scale=0.268]{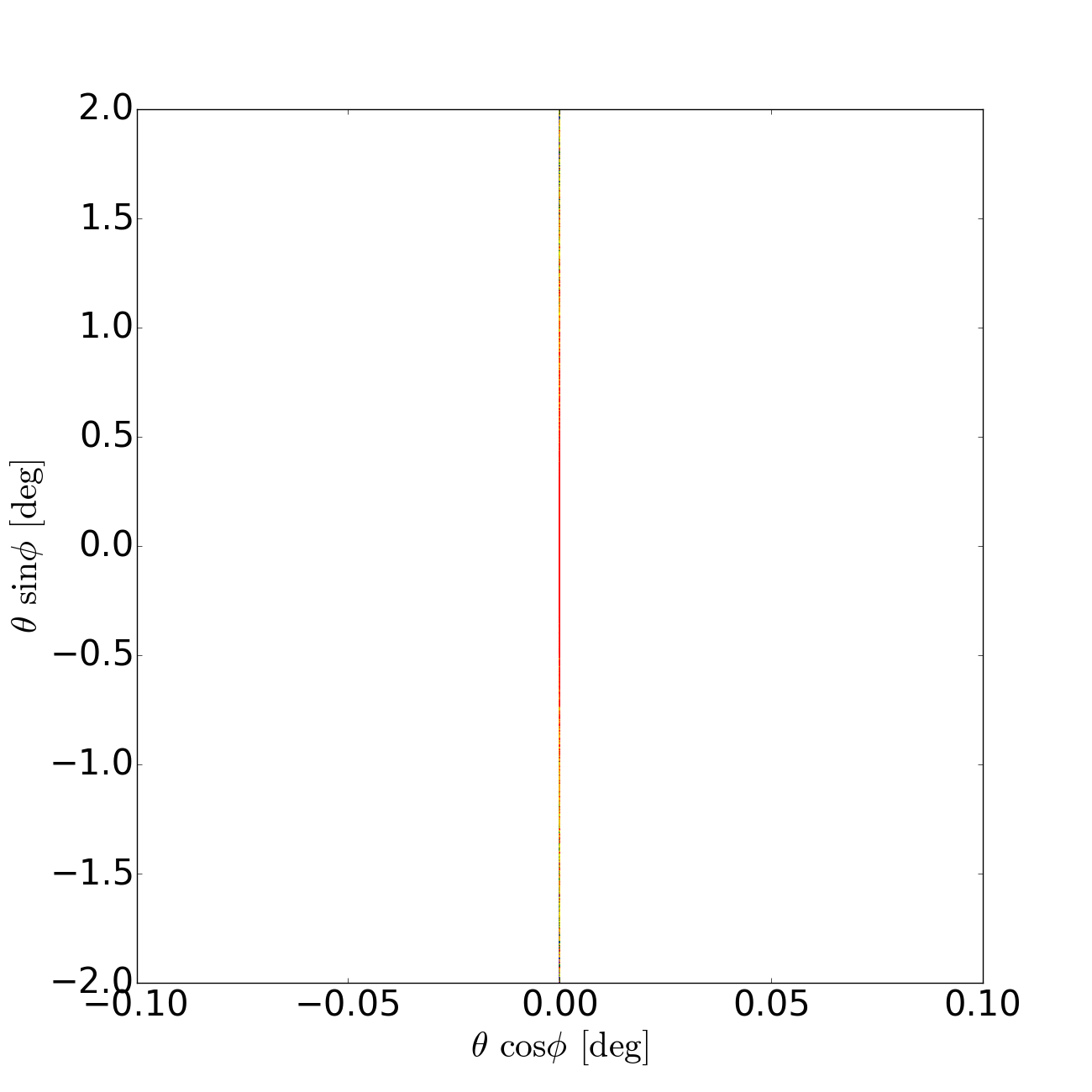}
\includegraphics[scale=0.268]{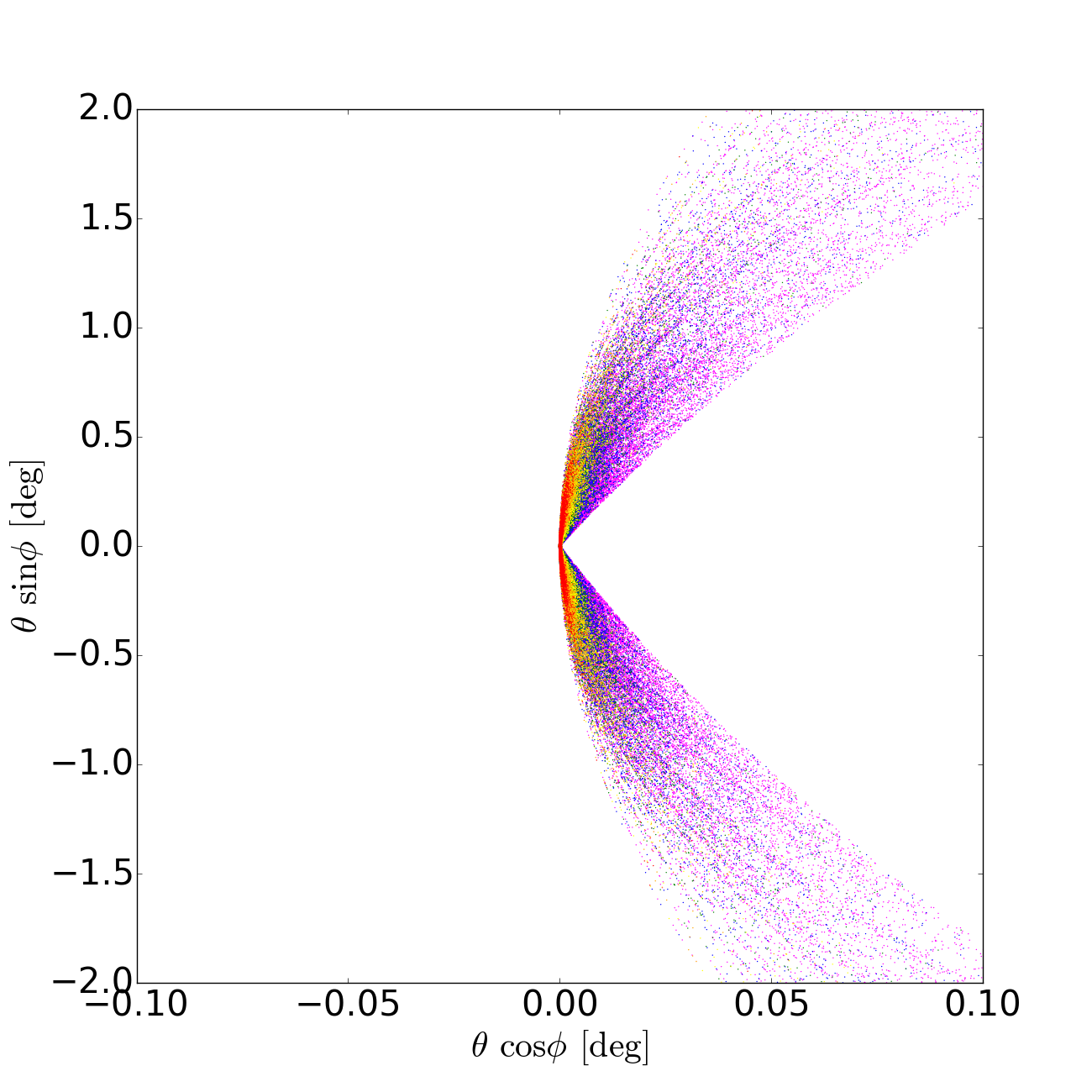}
\includegraphics[scale=0.268]{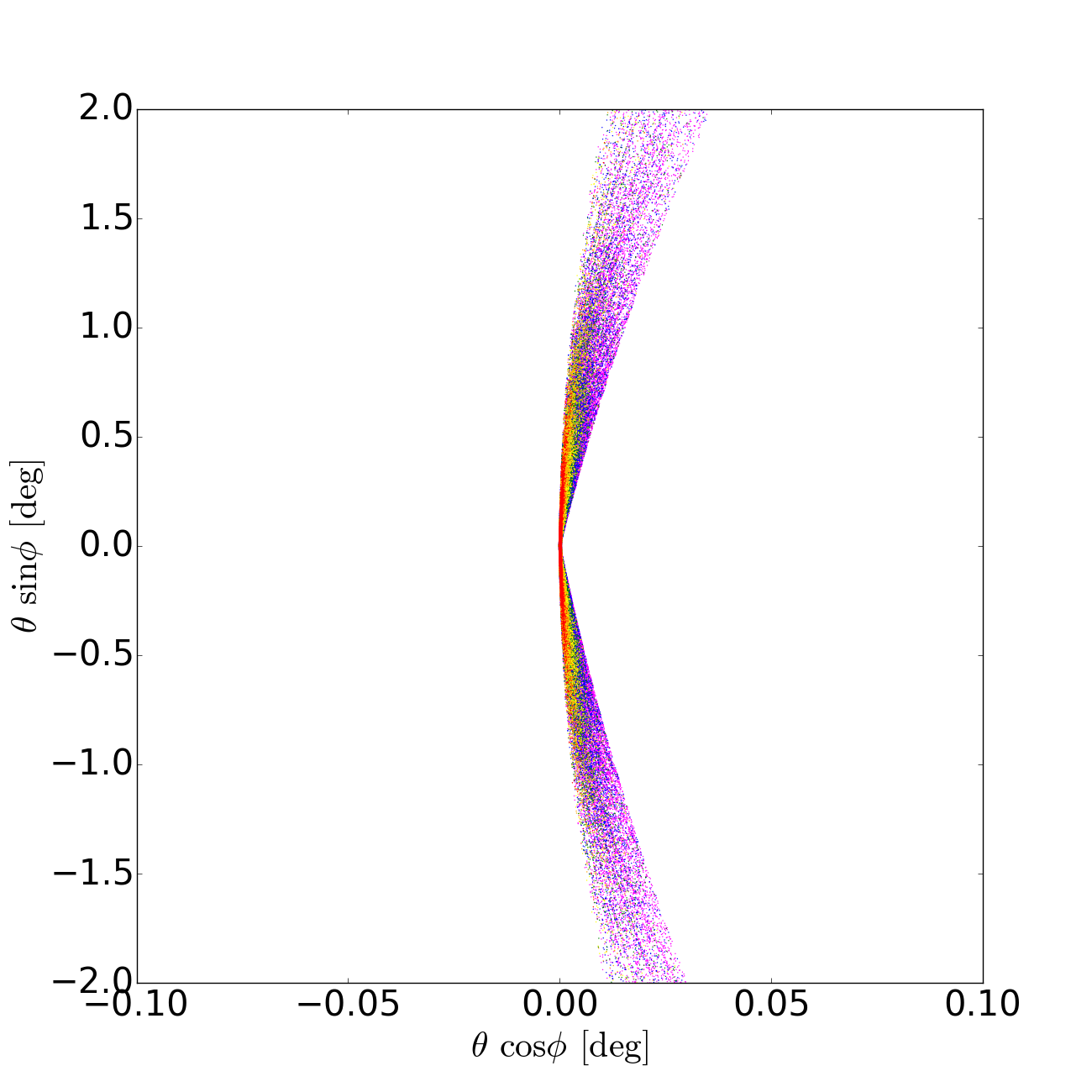}
\caption{Energy-dependent sky maps for a uniform magnetic field with $B = 10^{-15}\,{\rm G}$. We show the cases of a tightly collimated jet with magnetic field parallel (top left), perpendicular (top right), and tilted by $45\,\deg$ (bottom left) and $75\,\deg$ (bottom right) to the blazar jet direction which is taken to be along the line of sight.
The different colors represent the following energy ranges: $5 \-- 10\,{\rm GeV}$ (magenta), $10 \-- 15\,{\rm GeV}$ (blue), $15 \-- 20\,{\rm GeV}$ (green), $20 \-- 30\,{\rm GeV}$ (yellow), $30 \-- 50\,{\rm GeV}$ (orange), $50 \-- 100\,{\rm GeV}$ (red).}
\label{fig:HomField}
\end{figure*}

In Fig.~\ref{fig:HomField} these different cases are shown for a magnetic field of strength $10^{-15} \, {\rm G}$, assuming that the gamma rays are emitted in a jet with a half-opening angle of $5^\circ$ and with energy $E_{\rm TeV} = 10 \, {\rm TeV}$.  The results for the three cases with a specific focus on their energy dependence are shown in Fig.~\ref{fig:HomField}.

The results for the first case, in which the magnetic field is parallel to the jet axis, are rather intuitive and are shown in the upper left panel of Fig.~\ref{fig:HomField}. One can see that there is only one possible arrival direction, face-on, {\it i.e.}~$\theta=0^\circ$, which means that only electrons created with momenta parallel to the magnetic field lines, and thus not influenced by the Lorentz force, can reach the observer. 
Any electron that deviates from the line of sight will have a trajectory that leaves the plane spanned by the
line of sight and the velocity direction of the initial TeV photon and will not reach the observer.

The second case, shown in the top right panel of Fig.~\ref{fig:HomField}, has a magnetic field perpendicular to the line of sight. 
Here only photons arriving in a plane perpendicular to the magnetic field are detected. This means that the parent-electrons of these photons describe circular motion in this same plane. If an electron has a velocity component \emph{parallel} to the magnetic field, it is
initially directed away from the line of sight, and there is no component of the Lorentz force that can bend it
back towards the observer.

In the case of an intermediate orientation of the magnetic field, illustrated through the bottom panel of Fig.~\ref{fig:HomField} for a tilt angle of $45^\circ$ (left) and $75^\circ$ (right), we obtain results between the two extreme cases previously discussed, as expected. It is interesting to notice that the patterns are now smeared out since electrons from a range of directions can be directed towards the observer. Still, the dilution of the signal is small compared to the actual deflection, and hence this can be observed. Therefore, relevant information can still be extracted from sky maps by using the morphology of the arrival directions. 

\begin{figure*}
\center
\includegraphics[scale=0.22]{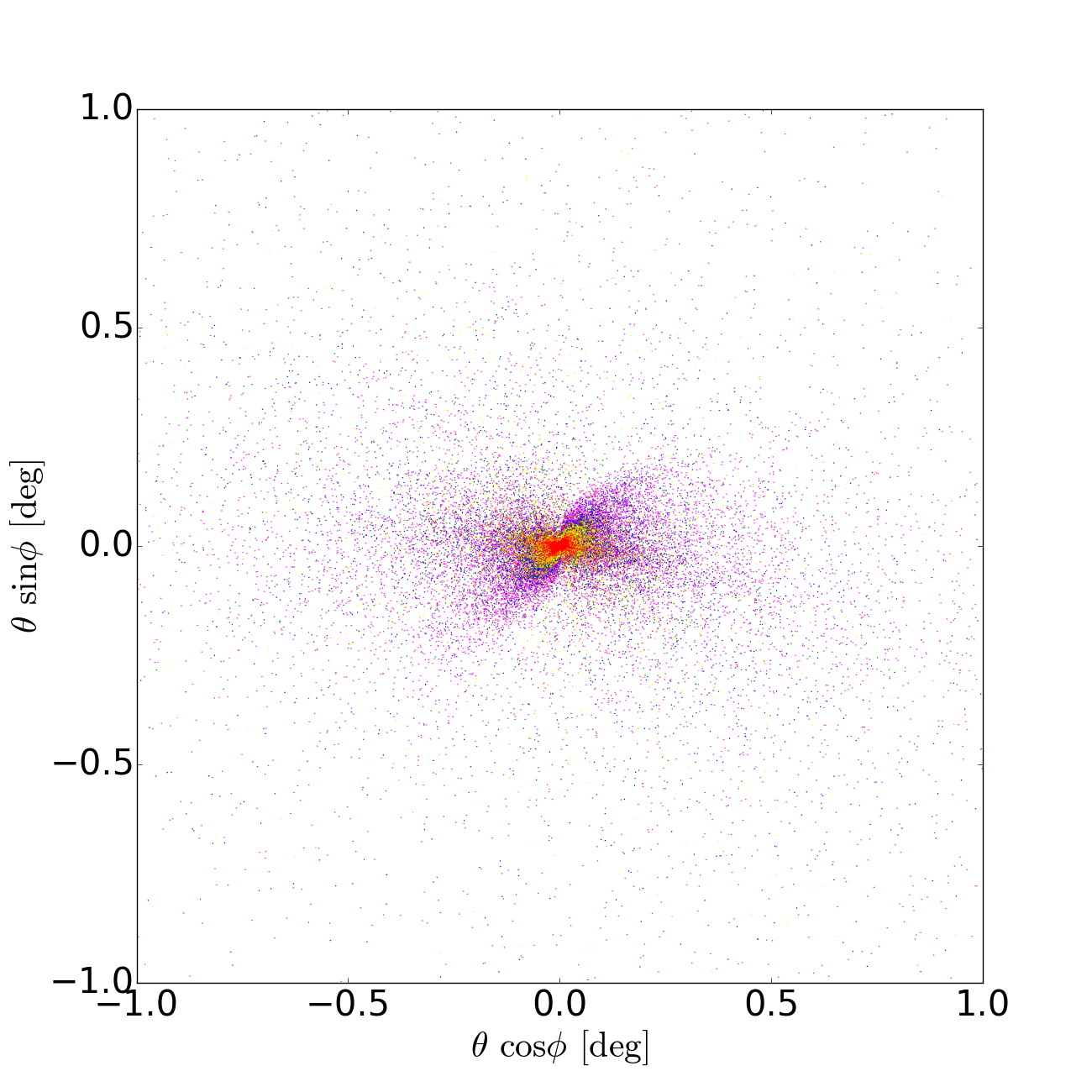}
\includegraphics[scale=0.22]{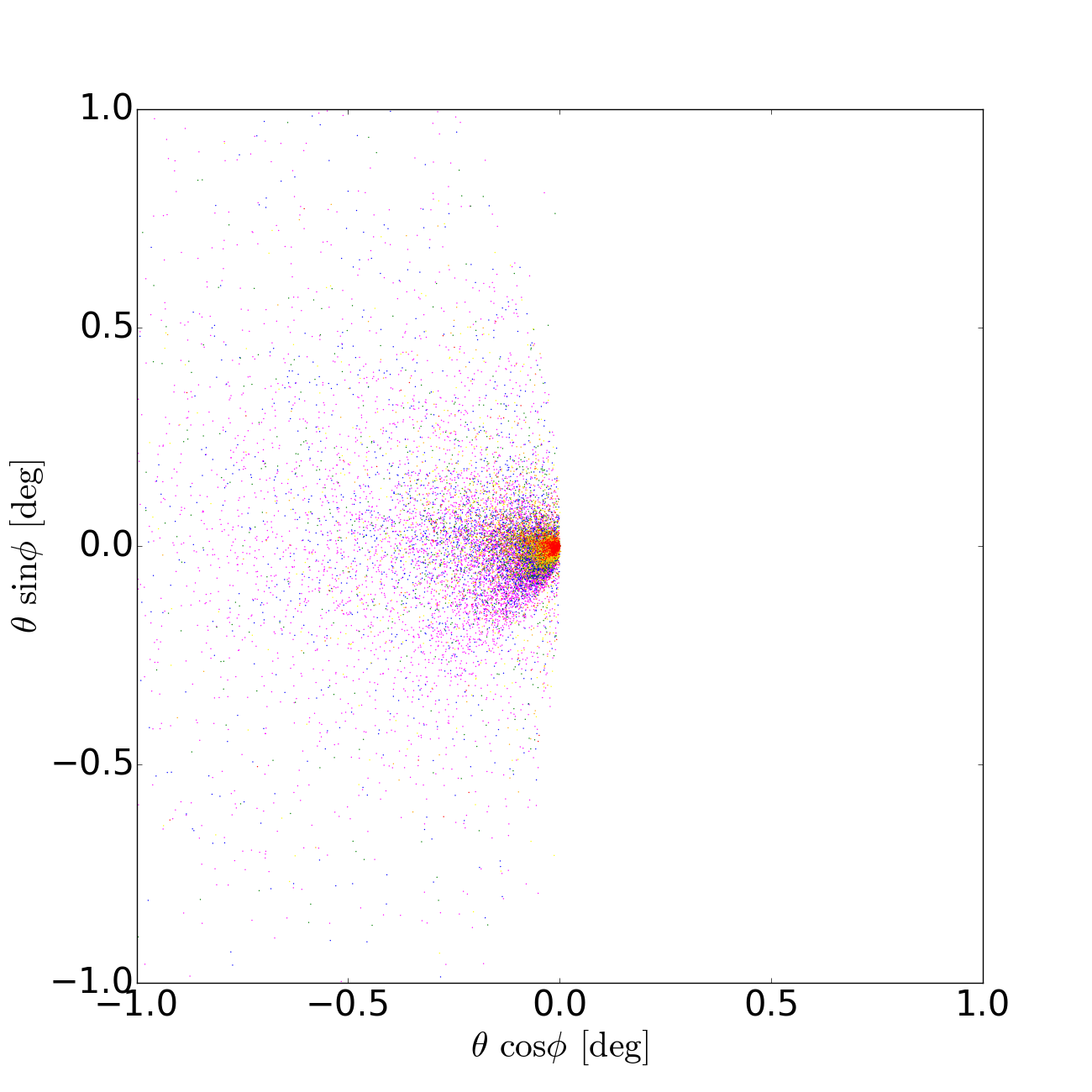}
\includegraphics[scale=0.22]{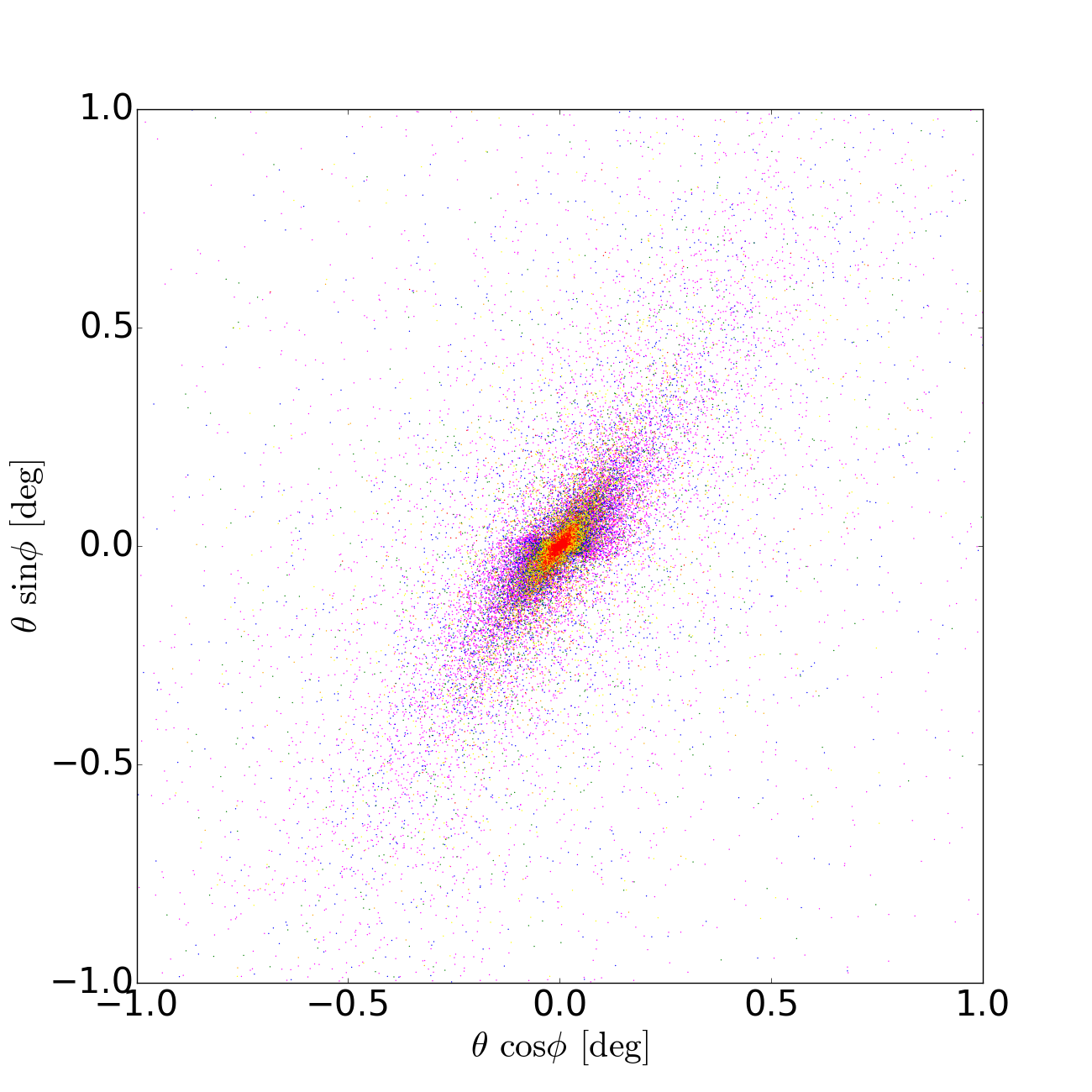}
\includegraphics[scale=0.22]{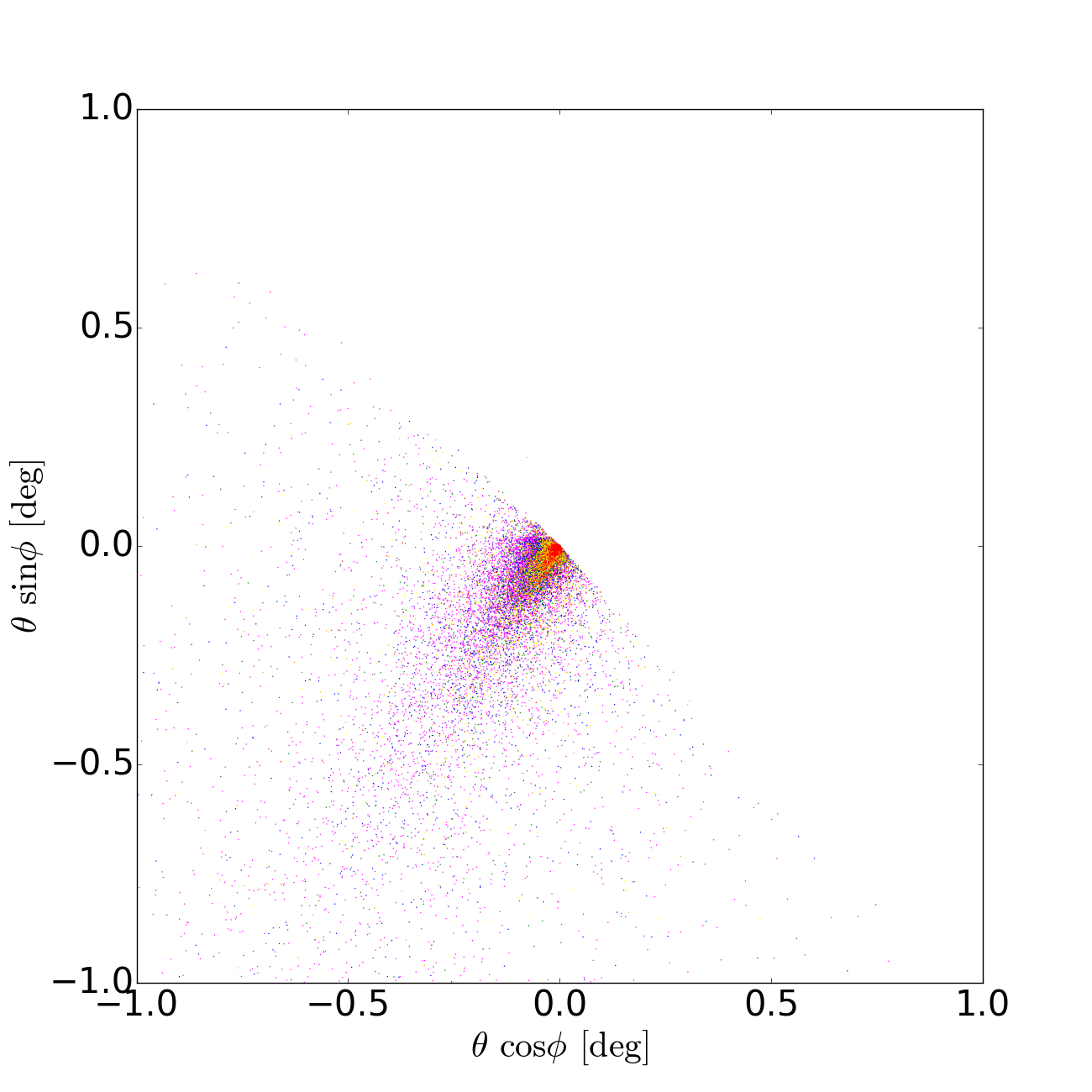}
\includegraphics[scale=0.22]{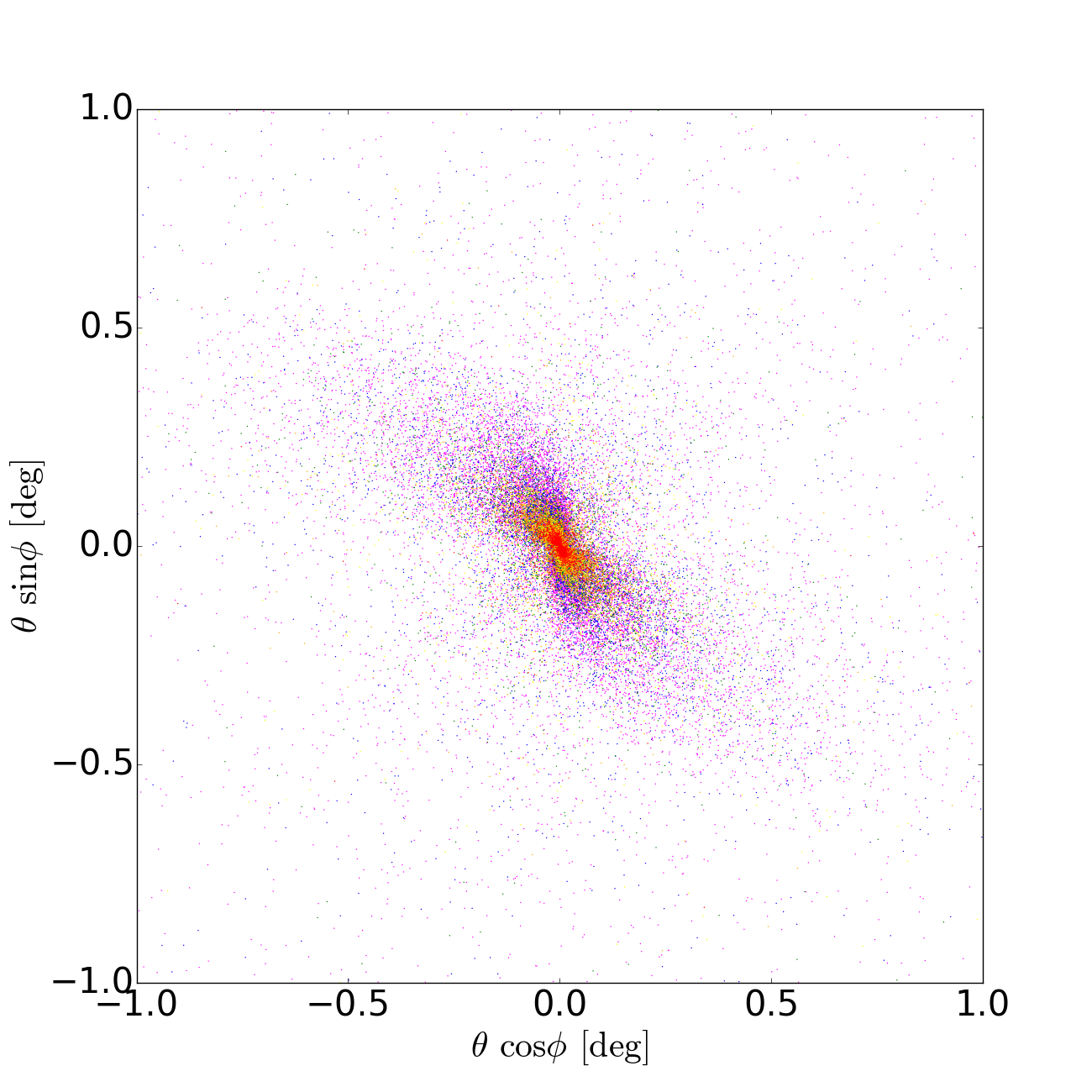}
\includegraphics[scale=0.22]{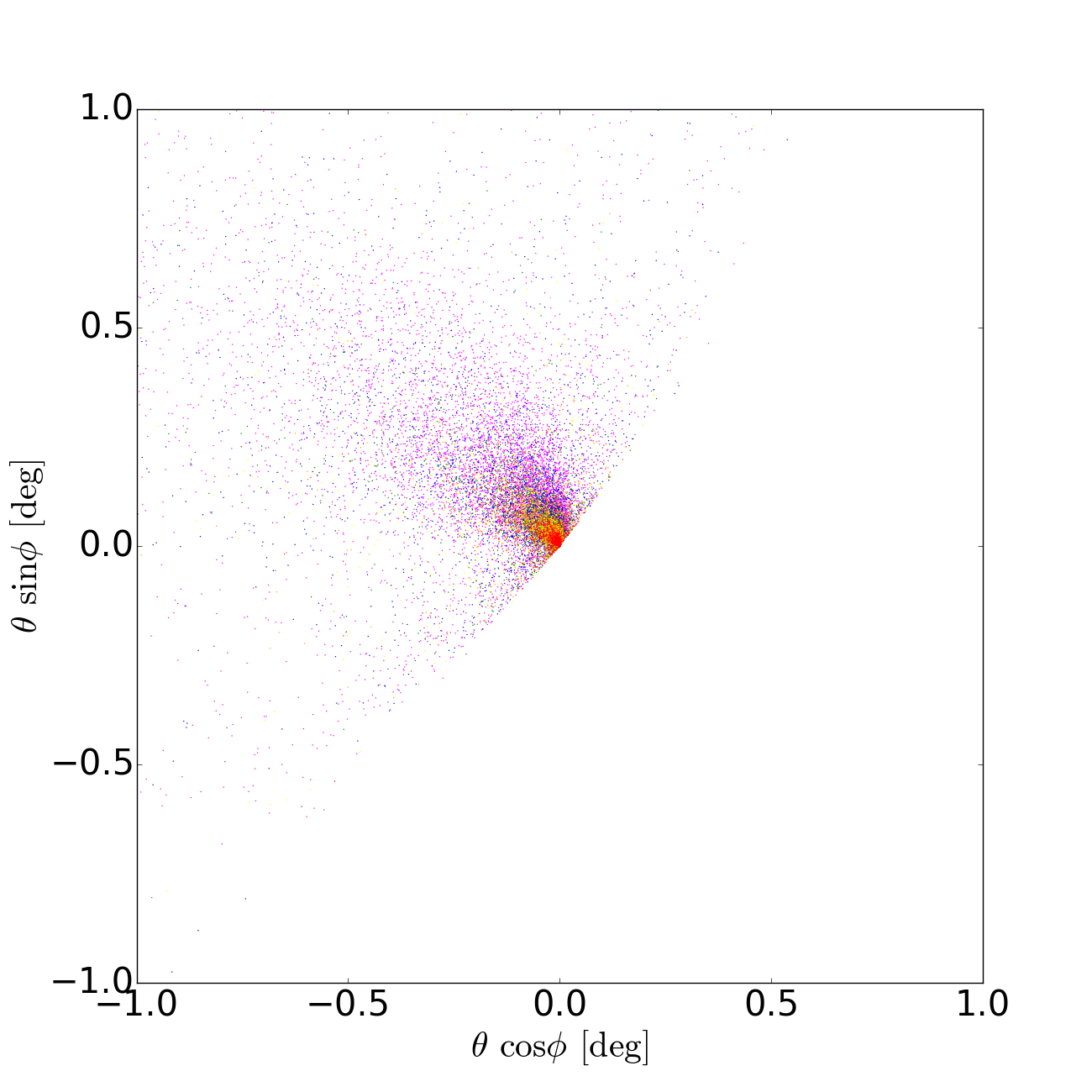}
\caption{Sky maps of arrival directions of photons from a blazar at a distance 
$D_{\rm s} = 1\,{\rm Gpc}$ emitting photons with energy 
$E_{\rm TeV} = 10\,{\rm TeV}$ in a jet with a half opening angle of $\Psi = 5^\circ$ directed at the 
observer (left column) and tilted by $5^\circ$ with respect to the line of sight (right column), respectively. 
The magnetic field is assumed to be stochastic with RMS strength of $B=10^{-15}\,{\rm G}$, coherence 
length $L_{\rm c} \simeq 120\,{\rm Mpc}$, and maximal negative (upper panels, $f_H=-1$),
null (central, $f_H=0$) and maximal 
positive (lower panels, $f_H=+1$) helicities, respectively. The colors represent the same energies as in 
Fig.~\ref{fig:HomField}.}
\label{fig:test5g}
\end{figure*}

\begin{figure*}[t]
\includegraphics[scale=0.179]{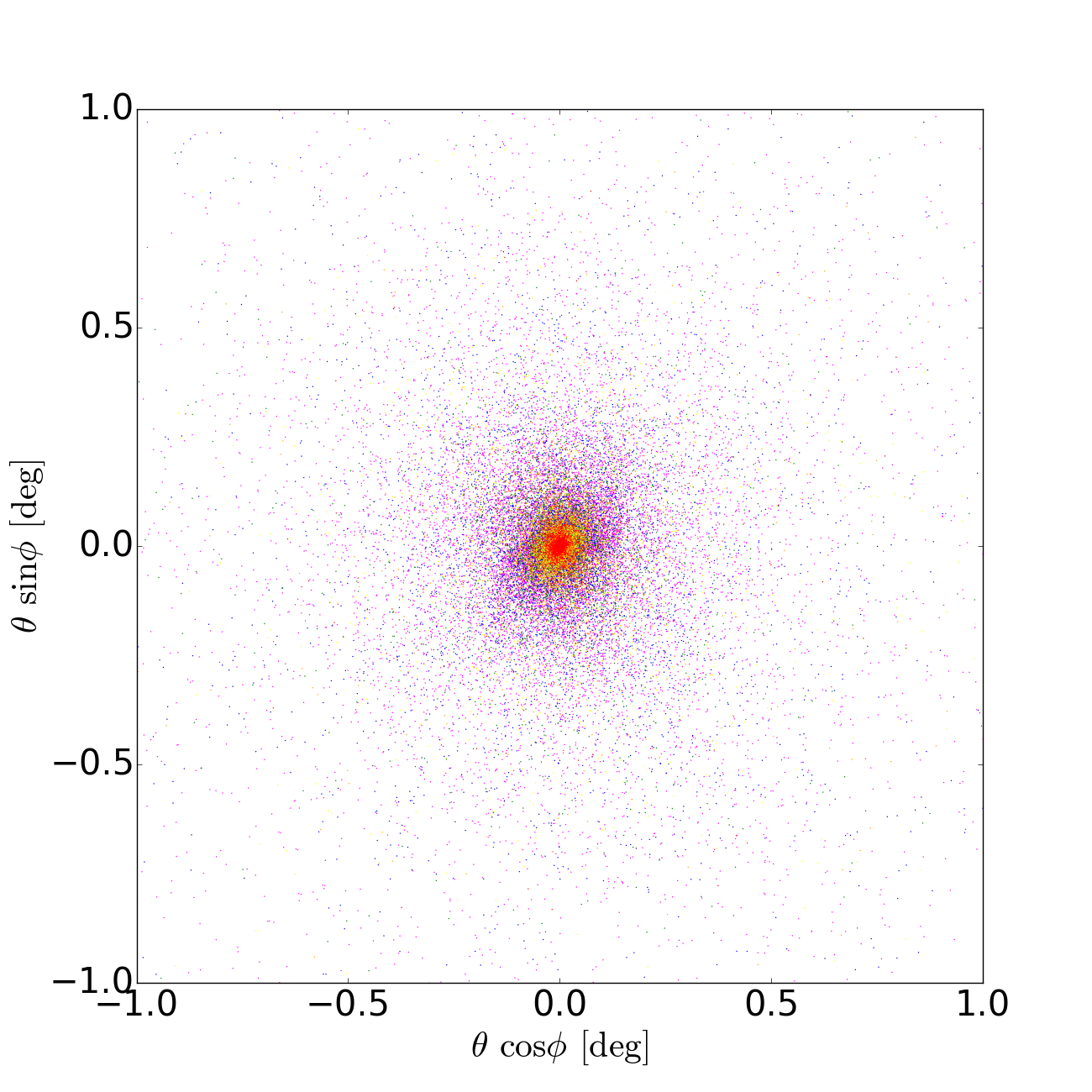}
\includegraphics[scale=0.179]{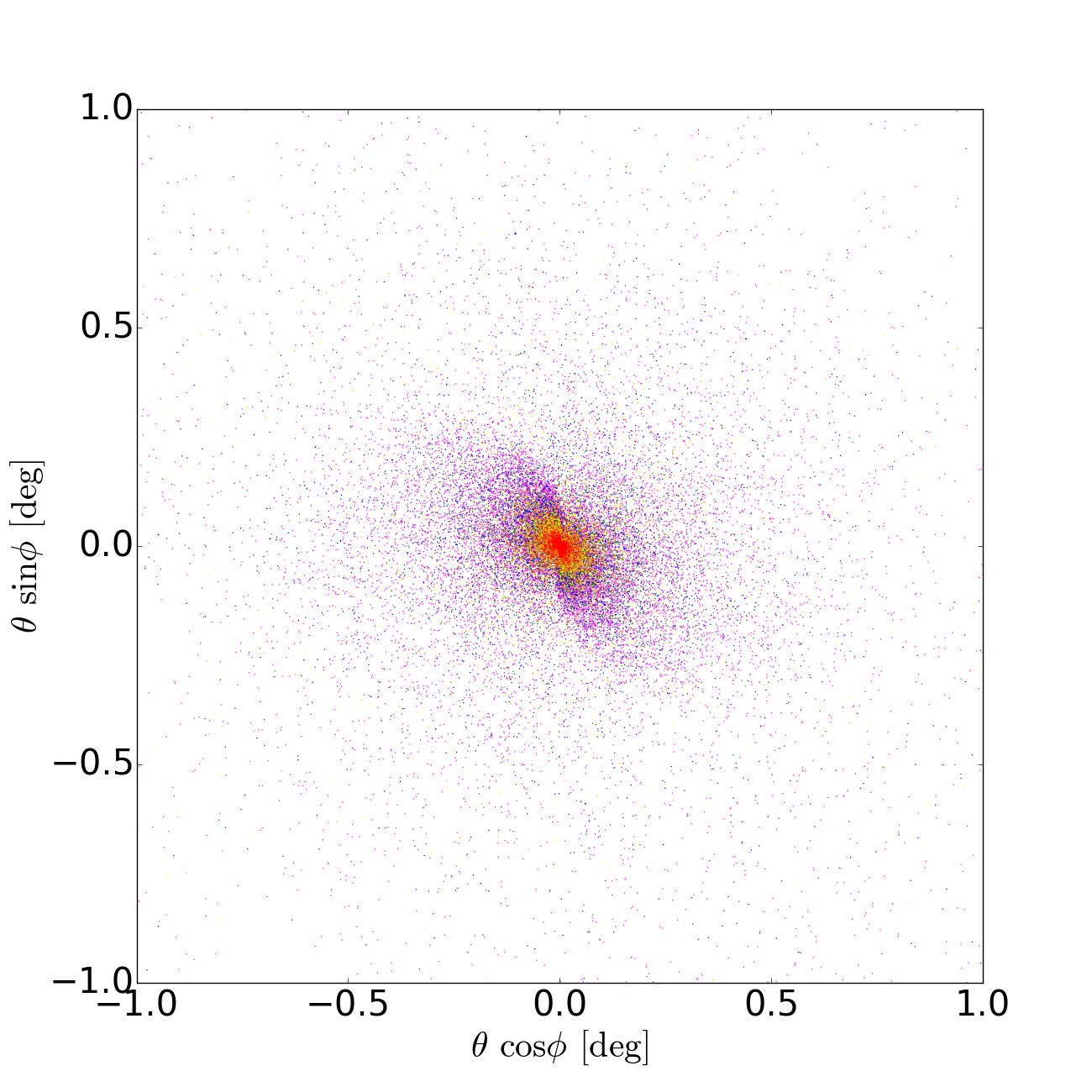}
\includegraphics[scale=0.179]{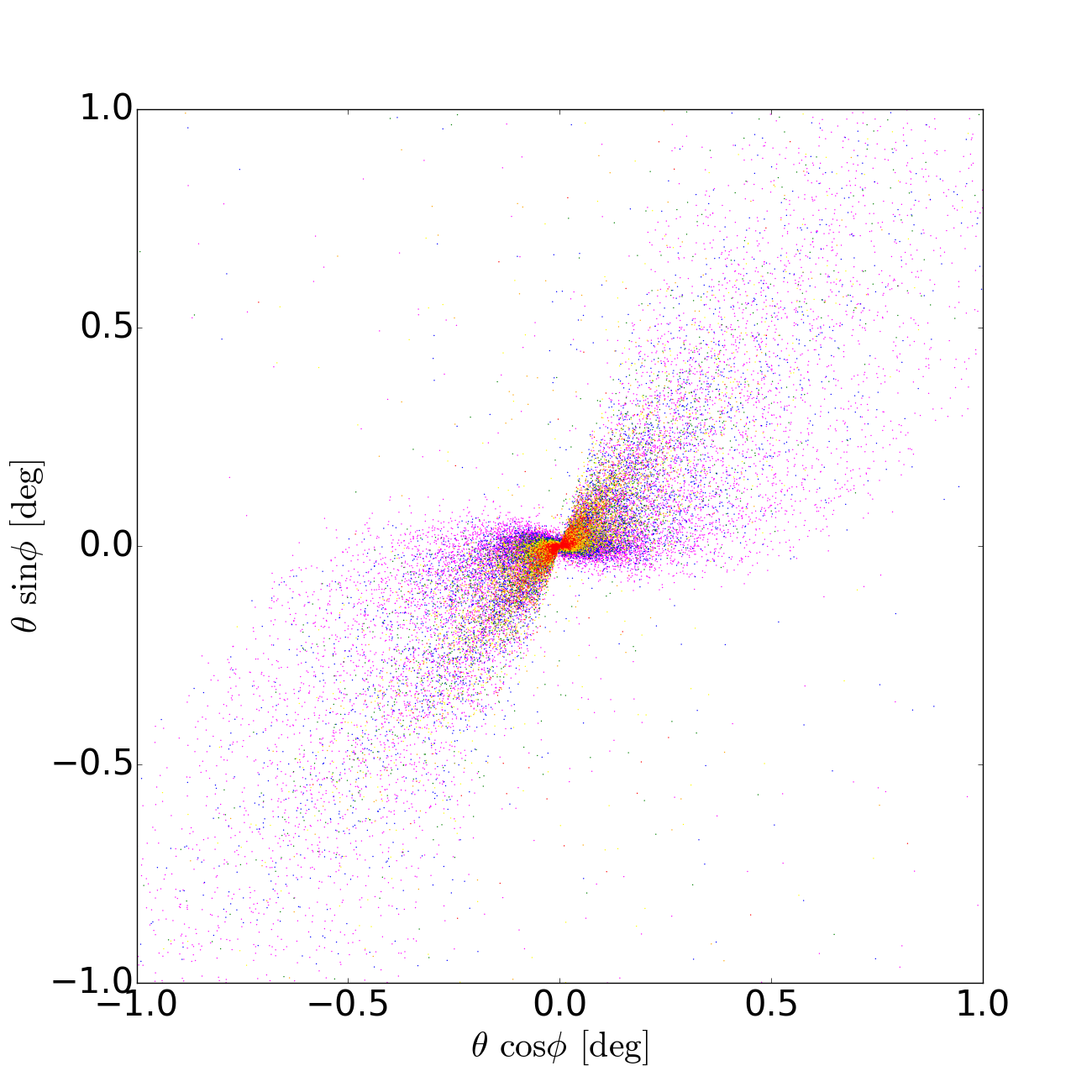}
\caption{Sky maps of arrival directions of photons from a blazar at a distance $D_{\rm s} = 1\,{\rm Gpc}$ emitting photons with energy 
$E_{\rm TeV} = 10\,{\rm TeV}$ in a jet with a half opening angle of $\Psi = 5^\circ$ directed at the observer. The magnetic field is assumed to be stochastic 
with RMS strength of $B=10^{-15}\,{\rm G}$ and a coherence length of $L_{\rm c} \simeq 50\,{\rm Mpc}$ (left), $L_{\rm c} \simeq 150\,{\rm Mpc}$ (center) 
and $L_{\rm c} \simeq 250\,{\rm Mpc}$ (right) for $f_{H} = +1$. The colors represent the same energies as in Fig.~\ref{fig:HomField}.}
\label{fig:diffLc}
\end{figure*}

\subsection{Stochastic Helical Magnetic Fields} \label{sec:HelMag1}

Now we introduce magnetic helicity to the simulations. The source is assumed to have a half opening angle 
$\Psi = 5^\circ$. We take the field to be stochastic with a Batchelor spectrum as in Eq.~(\ref{EB120}).
As we are assuming the maximal helical case, {\it i.e.}~$f_{H} = \pm 1$, this also 
fixes the spectrum of $H_{B}(k)$ according to Eq.~(\ref{HBEB}). The field has an average field strength of $B_{\rm rms} = 10^{-15}\, {\rm G}$ and a correlation length of $L_{\rm c} \simeq 120\,{\rm Mpc}$.
Here, $B_{\rm rms}^{2}$ can be extracted from Eq.~(\ref{EB(k)}) by setting
\begin{equation}
\begin{split}
 B_{\rm rms}^{2} \equiv \langle \left| \mathbf{B}(\mathbf{x}) \right|^2) \rangle &= \frac{1}{(2 \pi)^{3}} \int \left| \tilde{B}(\mathbf{k}) \right|^2 \, {\rm d}^{3}k \\
 &= 8 \pi \int E_{B}(k) {\rm d}\ln k\,,
\end{split}
\end{equation}
while $L_{\rm c}$ is defined by \cite{Harari:2002dy}
\begin{equation} \label{Lcdef}
\begin{split}
L_c &= \frac{1}{(2\pi)^3}\frac{\pi}{B_{\rm rms}^2} \int | \tilde{B}(\mathbf{k}) |^2 k^{-1} \, {\rm d}^{3}k \\
&= \frac{8\pi^2}{B_{\rm rms}^2} \int E_B(k) k^{-1} {\rm d}\ln k \,, 
\end{split}
\end{equation}
such that for the $E_{B}$ defined in (\ref{EB120}) we have $L_{\rm c} \simeq 5 L_{\rm min}/8$, 
where $L_{\rm min}$ is the cutoff scale. 

We have simulated the propagation of gamma rays with initial energy $E_{\rm TeV} = 10 \, {\rm TeV}$ in the presence of stochastic magnetic fields with 
maximally negative ($f_H = -1$), zero ($f_H = 0$) and maximally positive ($f_H = +1$) helicities\footnote{We could generate the $f_H=-1$ gamma ray distribution by a parity reversal of the $f_H=+1$ plot. However, we 
simulate the two cases independently to show two different stochastic realizations.}.
To simulate $10^5$ photons in our standard scenario described above, {\it i.e.} with 
$D_{\rm s} = 1\,{\rm Gpc}$ and $B = 10^{-15}\,{\rm G}$, the current version of the code takes
$\sim 8$ hours on $64$ cores at $2300\,{\rm MHz}$.

\begin{figure*}
\center
\includegraphics[width=.67\columnwidth]{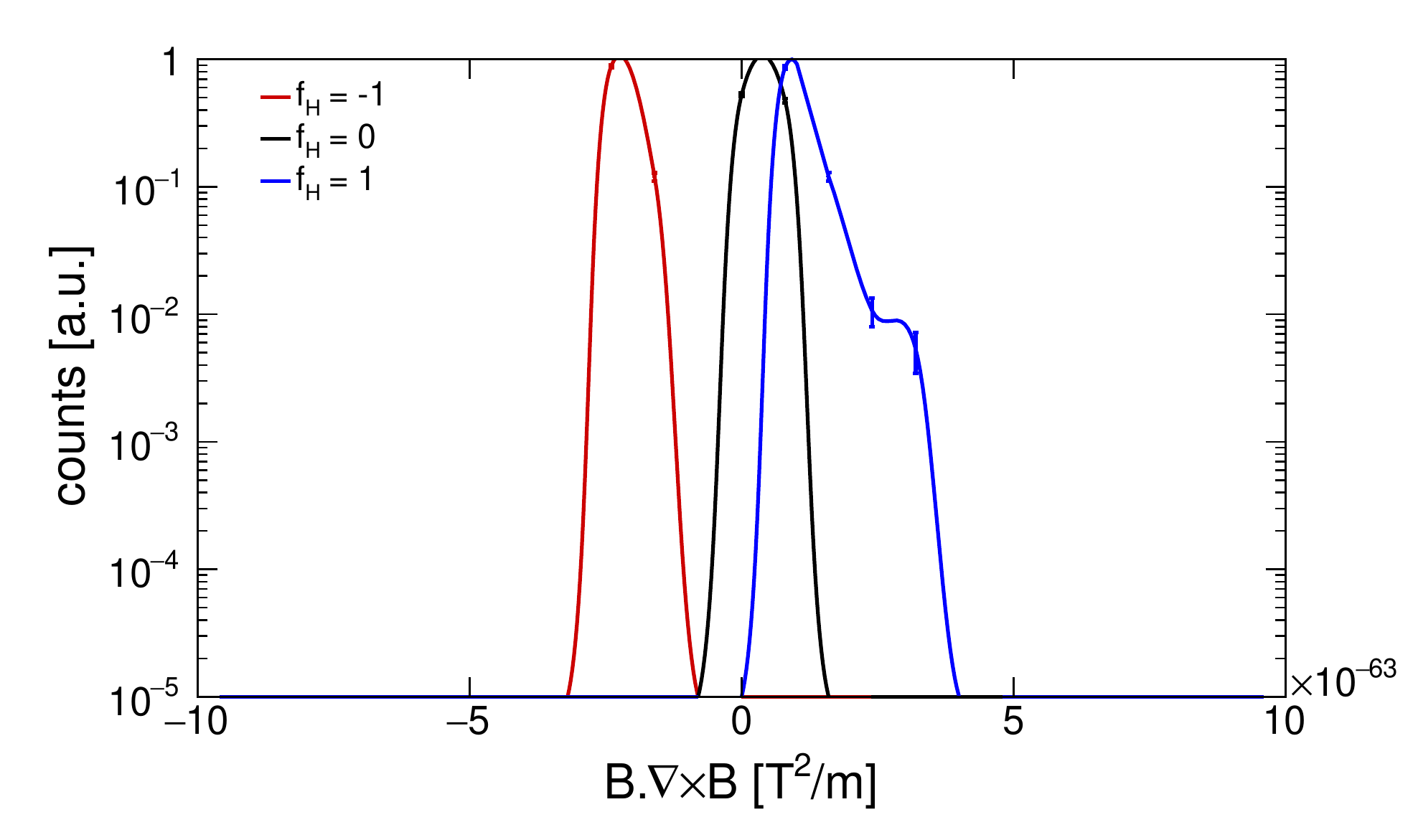}
\includegraphics[width=.67\columnwidth]{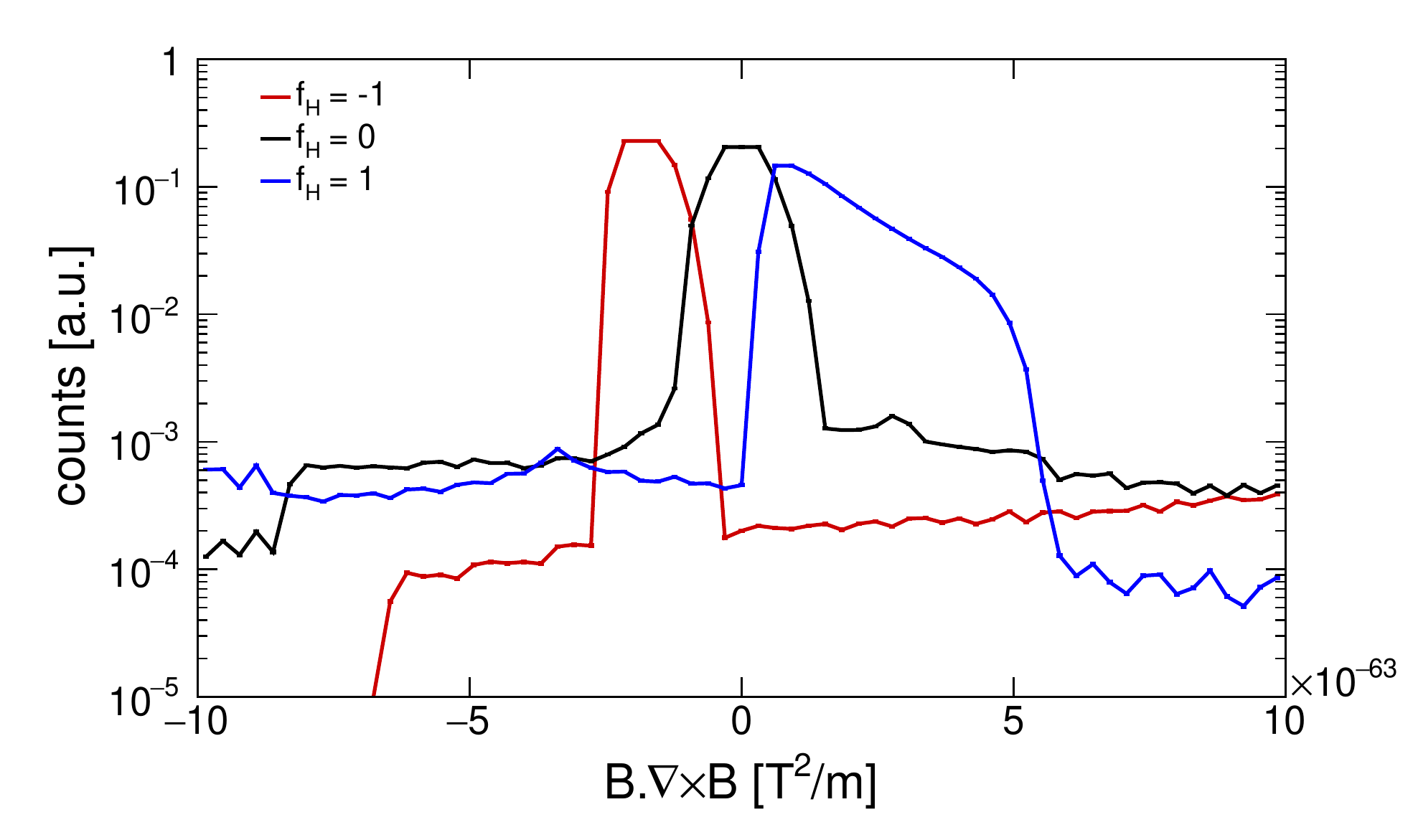}
\includegraphics[width=.67\columnwidth]{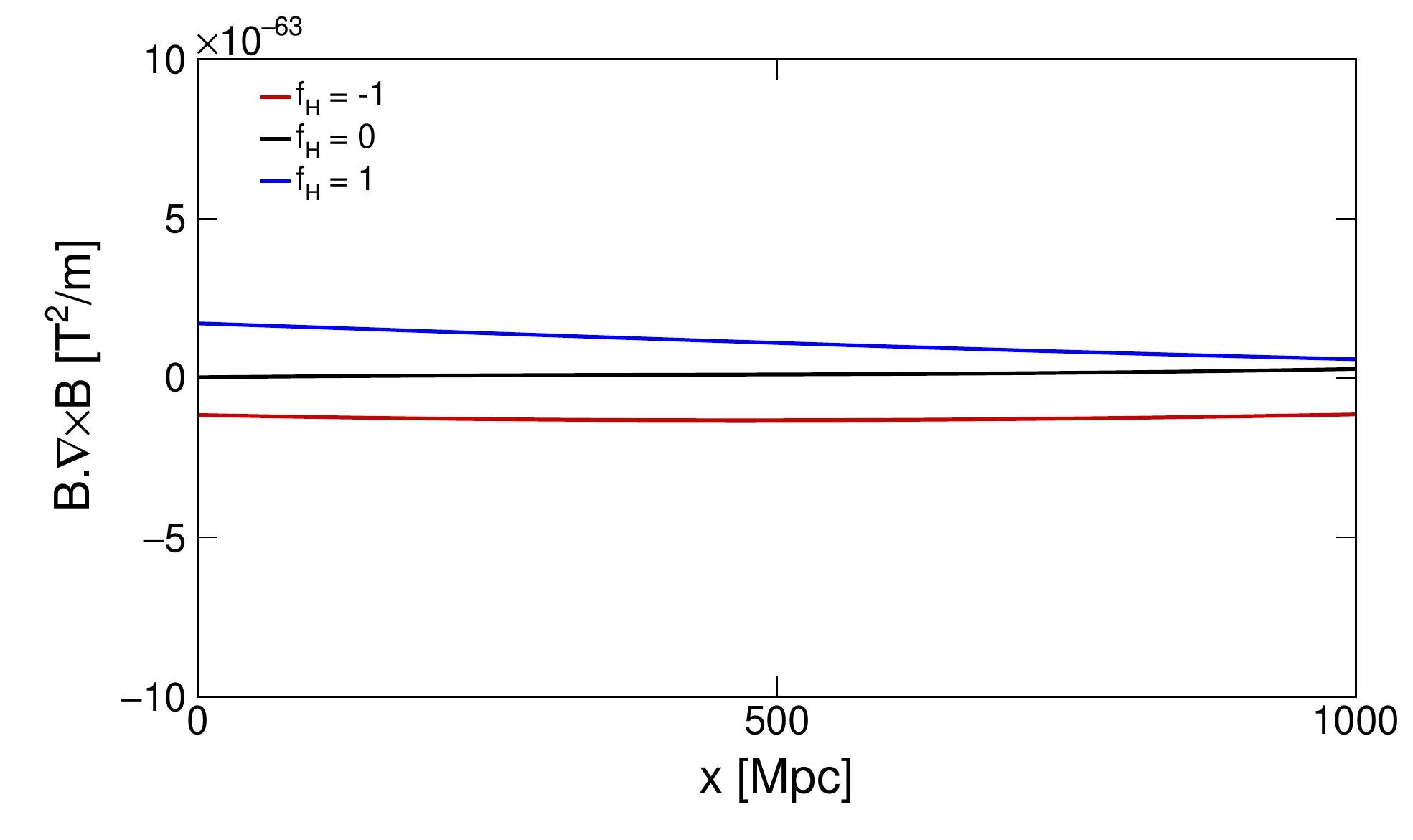}
\caption{Different magnetic helicity measures for the three cases shown in Fig.~\ref{fig:test5g}, {\it i.e.}~negative 
helicity ($f_{\rm H} = -1$, red), zero 
helicity ($f_{\rm H} = 0$, black) and positive helicity ($f_{\rm H} = +1$, blue). The left panel shows the total 
distribution of ``physical helicity'', defined as 
$\mathbf{B} \cdot \left( \nabla \times \mathbf{B} \right)$, in the whole simulation box, normalized to $1$. In the 
center panel the same measure is shown, 
however restricted only to the line of sight and the neighboring cells. Finally, the right panel shows the 
helicity values along the line of sight from the 
source (at $x = 0\,{\rm Mpc}$) to the observer (at $x = 1000\,{\rm Mpc}$).
}
\label{fig:helgrid}
\end{figure*}

The actual values for helicities for the whole simulation box as well as along the line sight are shown in Fig.~\ref{fig:helgrid} in order to illustrate to 
which extent statistics play a role. As one can see, both for the whole grid as well as just along the line of sight, which is more important to judge about 
the statistical significance for a given case, the helicity distribution corresponds to the sign it has been assigned. Furthermore, from the panel on the right, 
one can see that for these particular realizations, for $f_{\rm H} = +1$ the absolute magnitude of $\mathbf{B} \cdot \left( \nabla \times \mathbf{B} \right)$ 
is high close to the source and low close to the observer, while for $f_{\rm H} = -1$ it stays roughly equal along the propagation path. 
We can understand the qualitatively similar patterns for $f_H=\pm 1$, {\it i.e.} both patterns are spirals with similar
twist, by noting that pair production on average happens close to the source, and both cases have similar helicity 
measures in that region.
As a confirmation of this interpretation we found that in simulations in which the absolute value of $\mathbf{B} \cdot \left( \nabla \times \mathbf{B} \right)$ is small
close to the source, the spiral-like structures tend to be less distinct.

The sky map containing the arrival directions of gamma rays are shown in Fig.~\ref{fig:test5g}. We consider both the case for which the jet is directed along 
the line of sight (left column) and for which it is tilted by $5^\circ$ (right column). For the former one can see the impact of magnetic helicity by 
comparing the top ($f_H = -1$) and bottom ($f_H = +1$) panels. A remarkable spiral-like pattern is clearly visible, being left- or right-handed depending on 
whether the helicity is negative or positive, respectively. For zero helicity ($f_H = 0$, middle panels), on the other hand, no clear orientation can be seen. 

We show here the results for $L_{\rm c} \simeq 120 \, {\rm Mpc}$. For lower coherence lengths ($L_{\rm c} \lesssim 50\,{\rm Mpc}$) and $B \lesssim 10^{-15} {\rm G}$ we find that the arrival direction pattern is washed out, and it is not possible to infer the presence of helicity, 
thus confirming the analytical predictions of Ref.~\cite{Long:2015bda} for this combination of parameters using simulations. 
This can be seen in Fig.~\ref{fig:diffLc} where the results for different 
$L_{\rm c}$ and $f_{H} = +1$ are shown. While for $L_{\rm c} = 250\,{\rm Mpc}$ a clear characteristic spiral in the arrival directions can be seen, it 
becomes less visible for $L_{\rm c} = 150\,{\rm Mpc}$ and disappears for $L_{\rm c} = 50\,{\rm Mpc}$. 
Therefore, $L_{\rm c} = 120\,{\rm Mpc}$ 
is a reasonable choice in order to show the effects of helicity discussed below. It is also a valid value in certain 
magnetogenesis scenarios~\cite{DuNe}.

To understand the dependence of the spiral pattern on the coherence scale, we note that, for small coherence lengths, the spirals become too tight to be 
resolved, {\it i.e.}~their angular size becomes too small compared to the overall halo \cite{Long:2015bda}. It seems, however, that the quality of the
spiral might be highly sensitive to the specific values of the parameters of the setting such as $B$, $D_{\rm s}$ and $L_{\rm c}$ which we will further 
investigate in the future.

On the other hand, for larger coherence lengths the spirals tend to a straight line, similarly to the top right 
panel of Fig.~\ref{fig:HomField}, approaching the case of a simple uniform magnetic field. This, again, is rather 
intuitive, since if $L_{\rm c} \gtrsim D_{\rm s}$, the stochastic magnetic field will effectively be uniform on the length scales in question.

On the right hand side of Fig.~\ref{fig:test5g} we show the same scenario described above, but this time the direction of the jet is tilted by $5^\circ$ with respect to the line of sight. As one can see in the figure, this reduces the effective area of arrival directions and also the symmetry of the pattern. In our example, for instance, one of the ``arms'' of the spiral pattern or a part of it is removed. This enables us to apply the $Q$-statistics~\cite{Tashiro:2013ita} (discussed below) to relate the helicity of the field with the arrival directions of gamma rays. It should be noted that all findings of this and the previous sections are in good agreement with the analytic predictions of Ref.~\cite{Long:2015bda}. 

\subsection{Computing the $Q$-factors} \label{sec:HelMag3}

One possibility to quantify the role of magnetic helicity is to use the $Q$-statistics, introduced in 
Refs.~\cite{PhysRevD.87.123527,Tashiro:2013ita,Tashiro:2014gfa}. The key elements here are the 
observed energies and the arrival directions of gamma rays at Earth. 
For sets of photons
with energies $E_{1}$, $E_{2}$ and $E_{3}$ with $E_{1} < E_{2} < E_{3}$, the $Q$-statistics is given by 
\cite{Tashiro:2013ita}
\begin{equation} \label{Qstat}
Q(E_{1},E_{2},E_{3},R) = \frac{1}{N_{3}} \sum_{j=1}^{N_{3}} \left[ \bm{\eta}_{1j}(R) \times \bm{\eta}_{2j}(R) \right] \cdot \mathbf{n}_{j}(E_{3})\, ,
\end{equation}
where  $\mathbf{n}_{j}(E_{a})$ is the arrival direction of the $j$-th photon with energy $E_{a}$,
$N_{a}$ is the total number of photons of energy $E_a$, and $\bm{\eta}_{aj}(R)$ is given by
\begin{equation}
\bm{\eta}_{aj}(R) \equiv \frac{1}{N_a} \sum_{i\in {\cal D}_{a}({\bf n}_j(E_{3}),R)} \mathbf{n}_{i}(E_{a})\,,
\end{equation}
where ${\cal D}_a({\bf n}_j(E_{3}),R)$ represents the set of photons of energy $E_a$ that are located in a disk of
radius $R$ centered on ${\bf n}_j(E_{3})$. Essentially, the $Q$-statistics is the average value of the triple product of
photon arrival vectors of energies $E_{1},E_{2},E_{3}$ that lie within an angle $R$ of the highest energy photon ($E_{3}$).

As has been shown in Refs.~\cite{PhysRevD.87.123527,Tashiro:2013ita,Tashiro:2014gfa}, the calculation of the 
parity-odd statistics, or $Q$-statistics, should enable us, depending on the sign and general shape of the 
$Q$-factors for different values of $E_{1}$, $E_{2}$, $E_{3}$, and $R$, to draw conclusions about the helicity of 
the intervening helical magnetic field. 

\begin{figure*}
\includegraphics[scale=0.209]{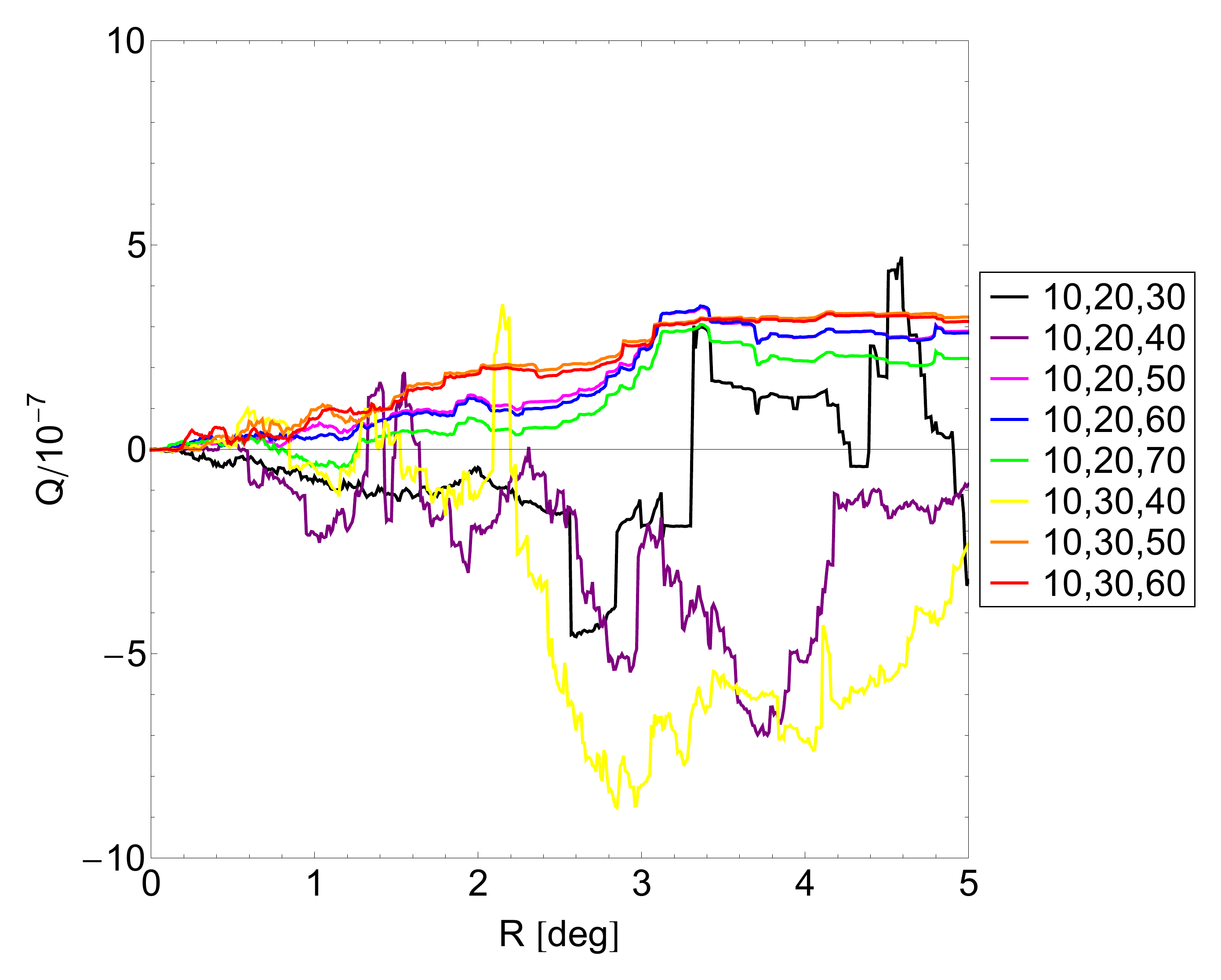}
\includegraphics[scale=0.209]{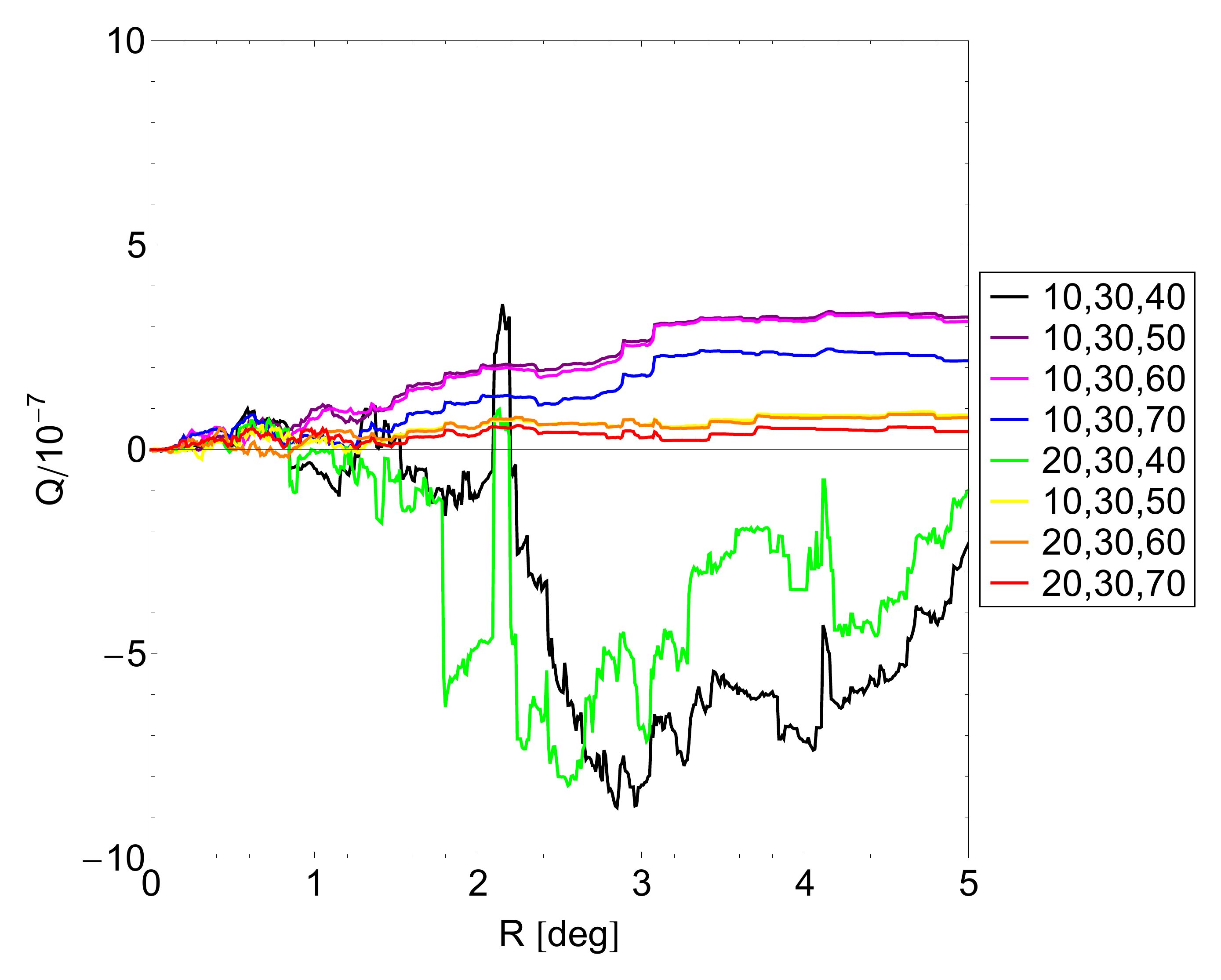}
\includegraphics[scale=0.209]{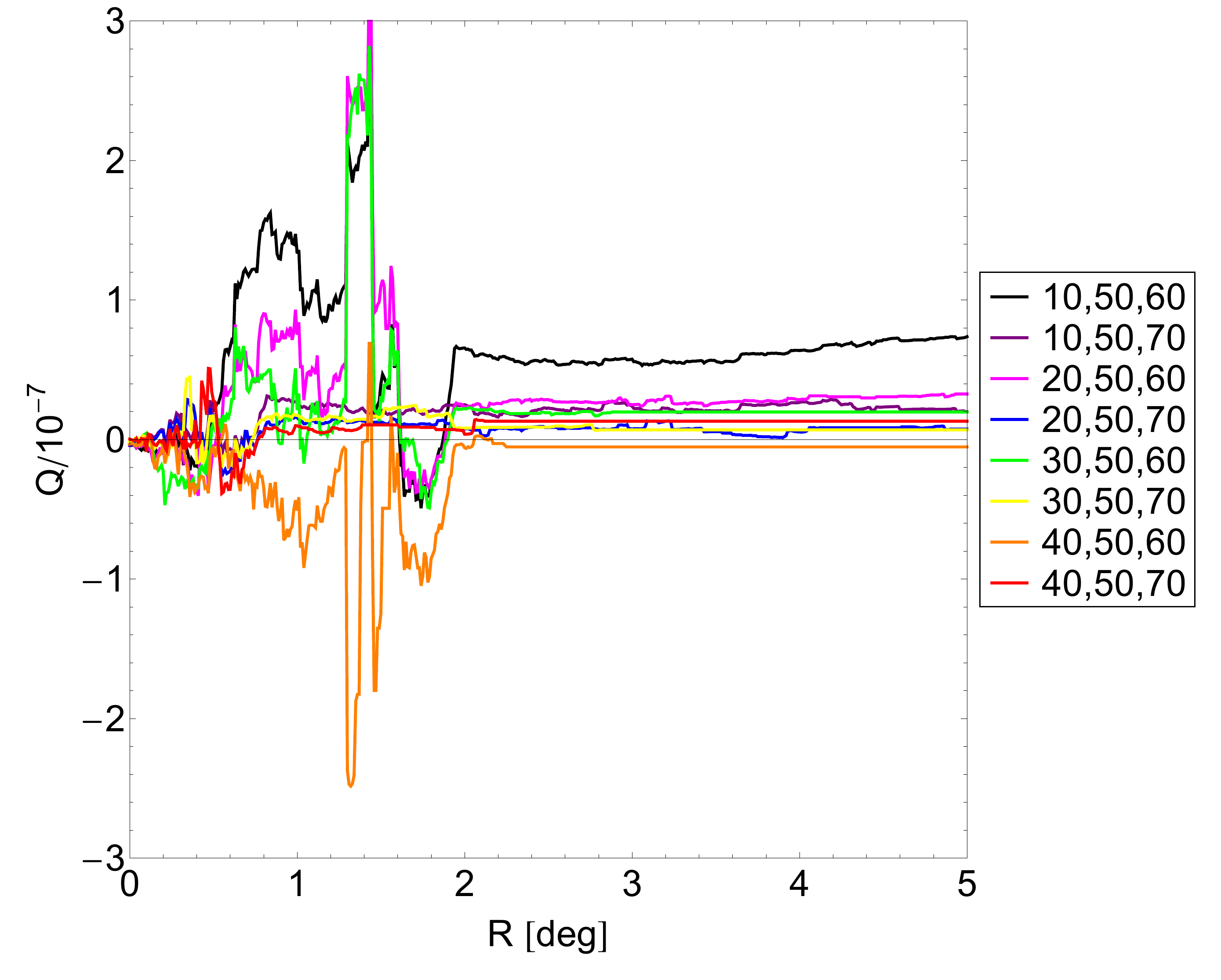}
\includegraphics[scale=0.209]{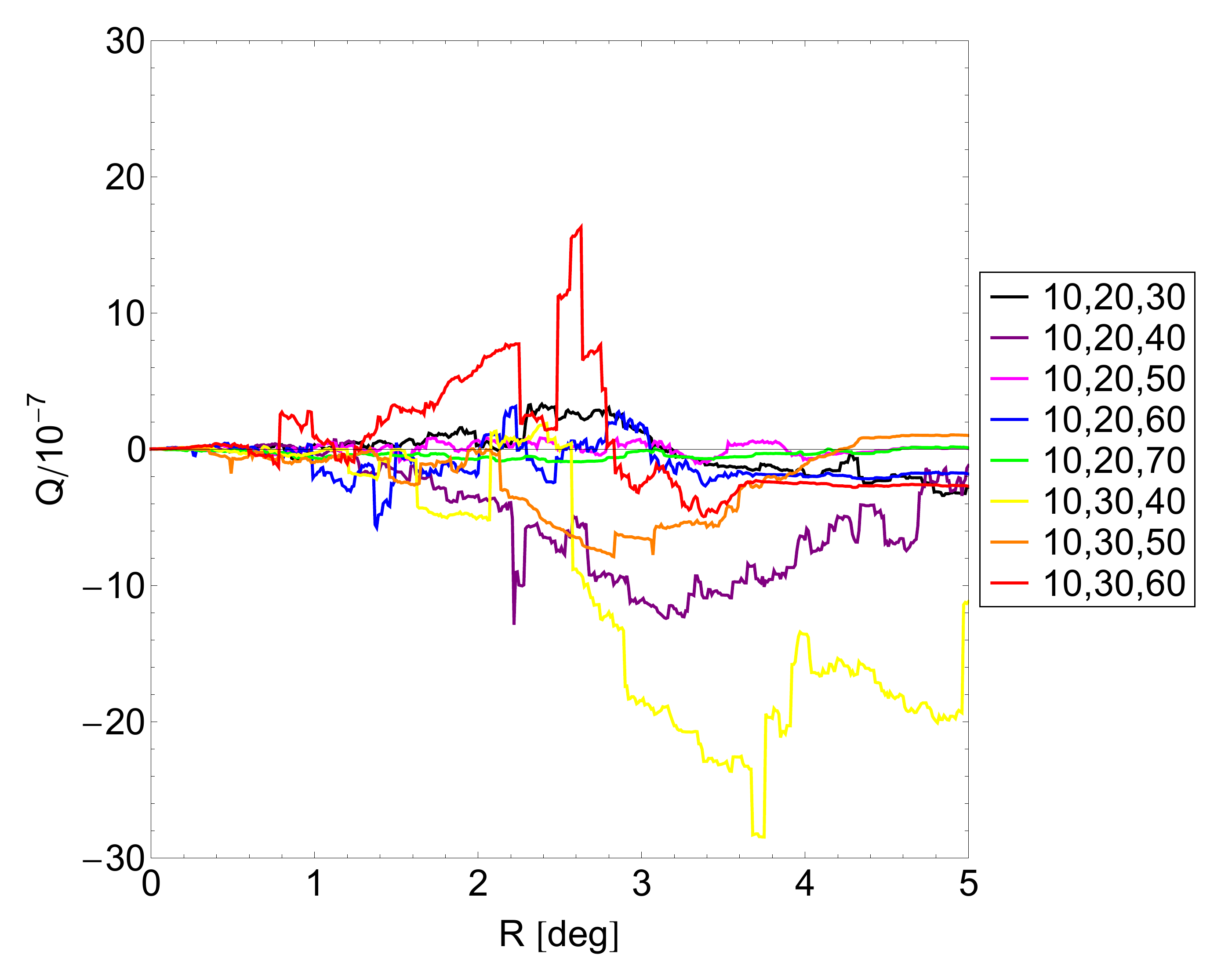}
\includegraphics[scale=0.209]{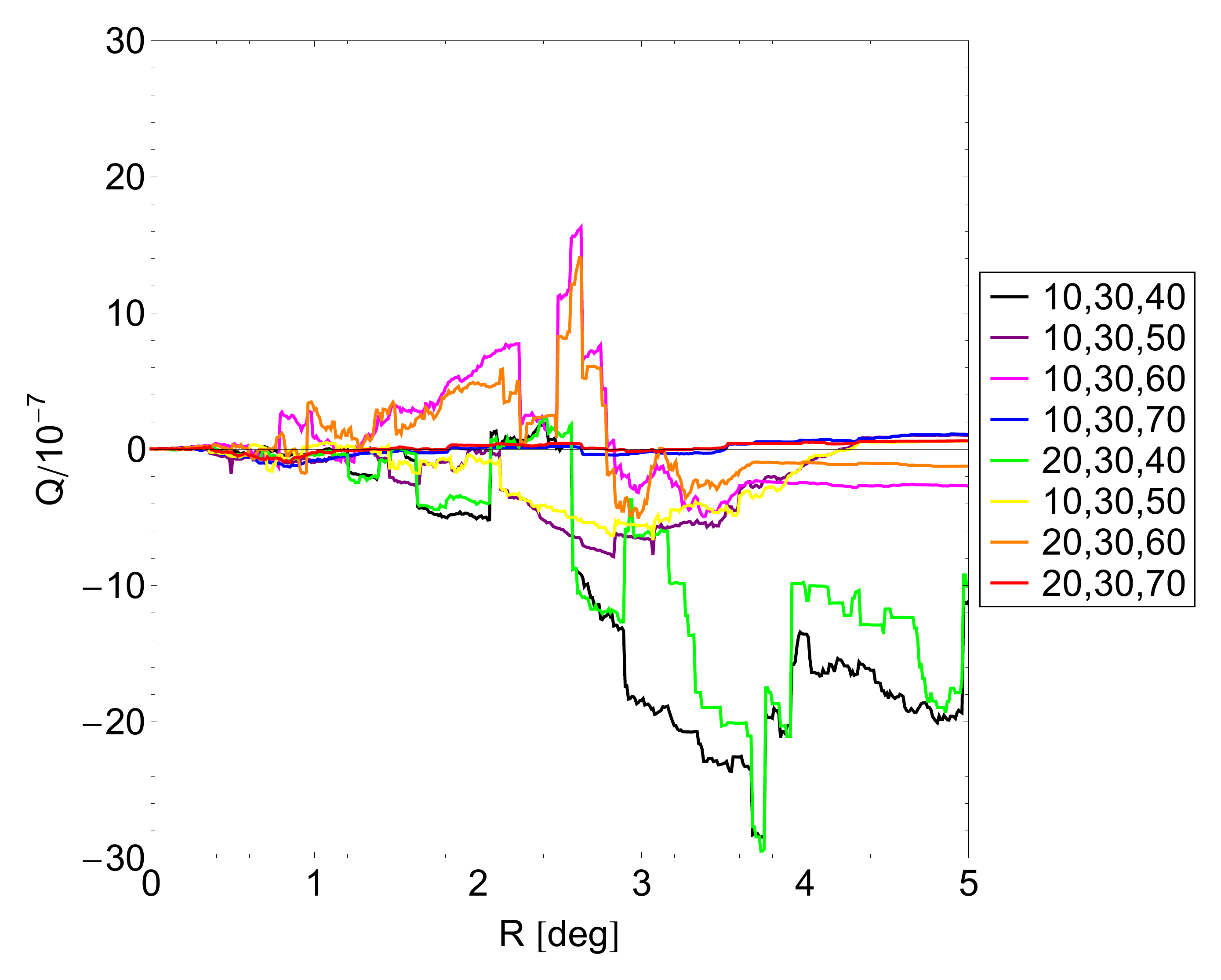}
\includegraphics[scale=0.209]{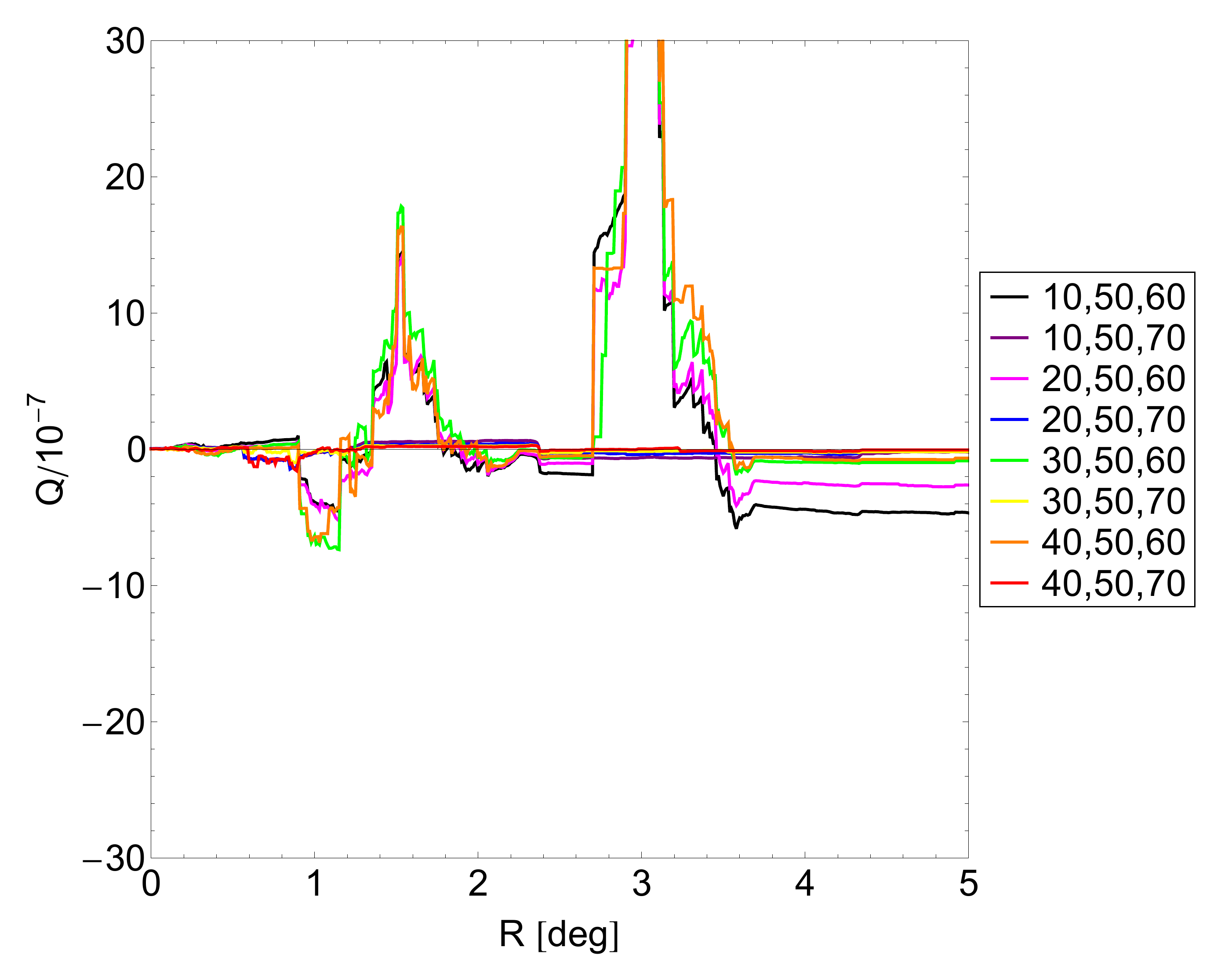}
\includegraphics[scale=0.209]{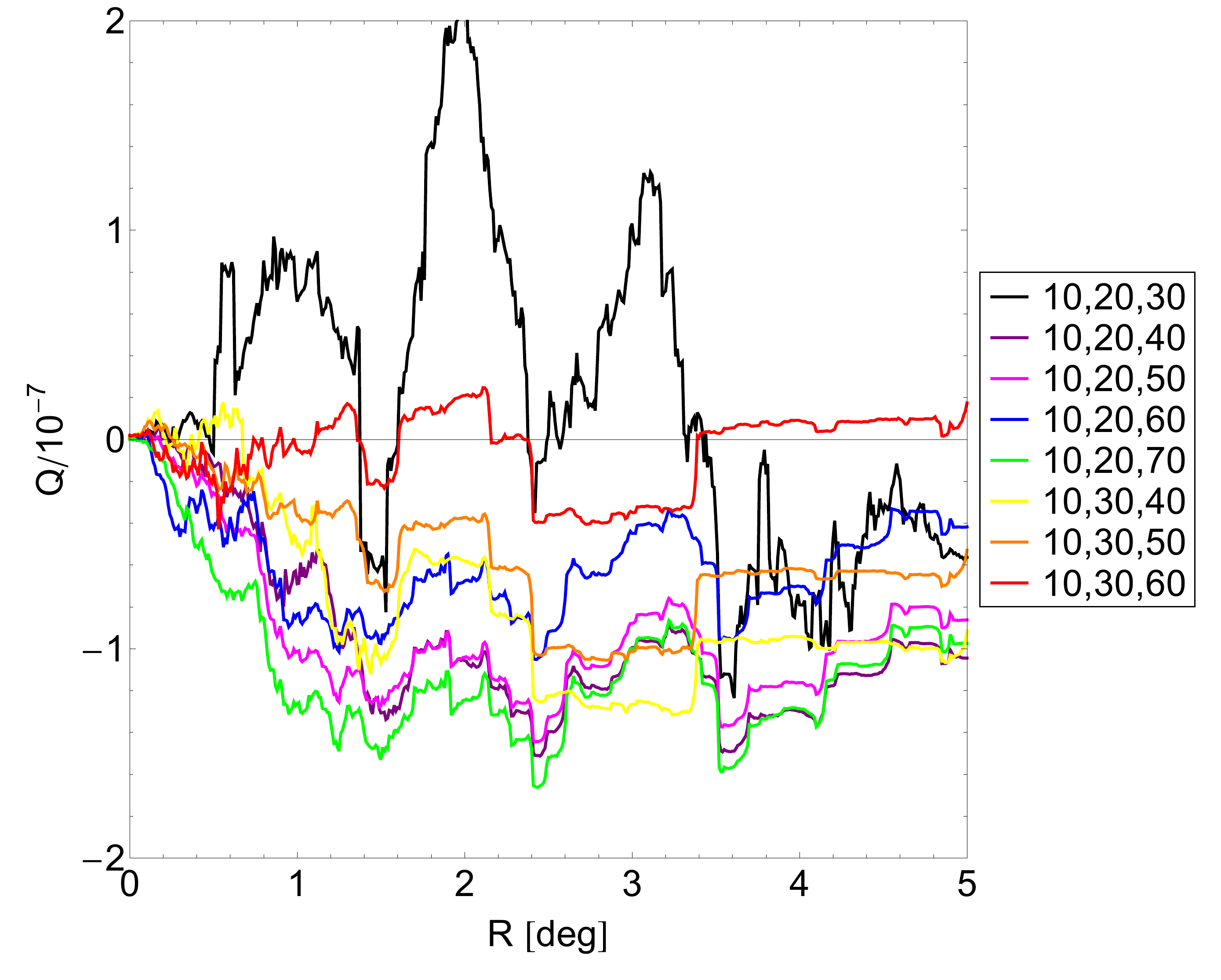}
\includegraphics[scale=0.209]{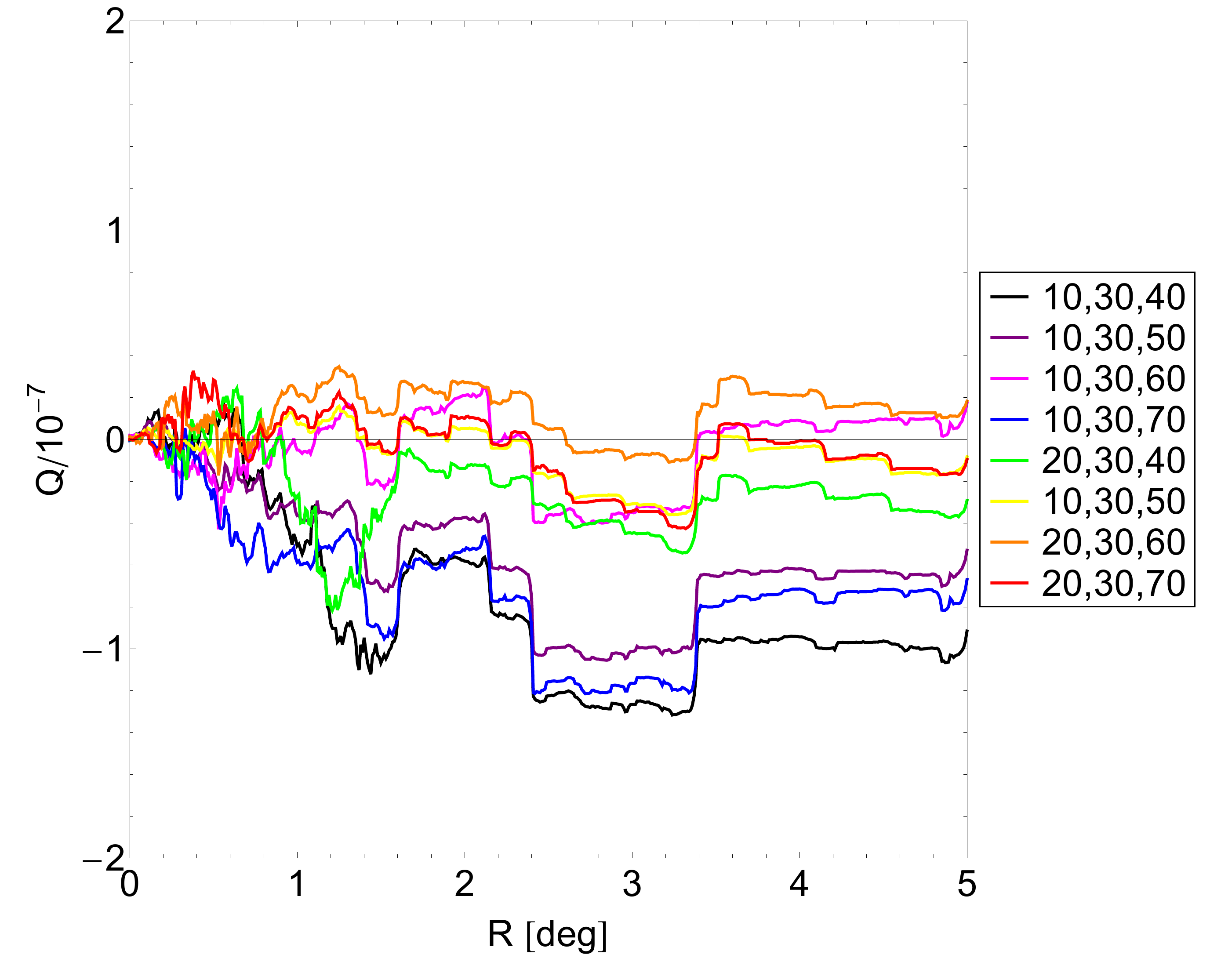}
\includegraphics[scale=0.209]{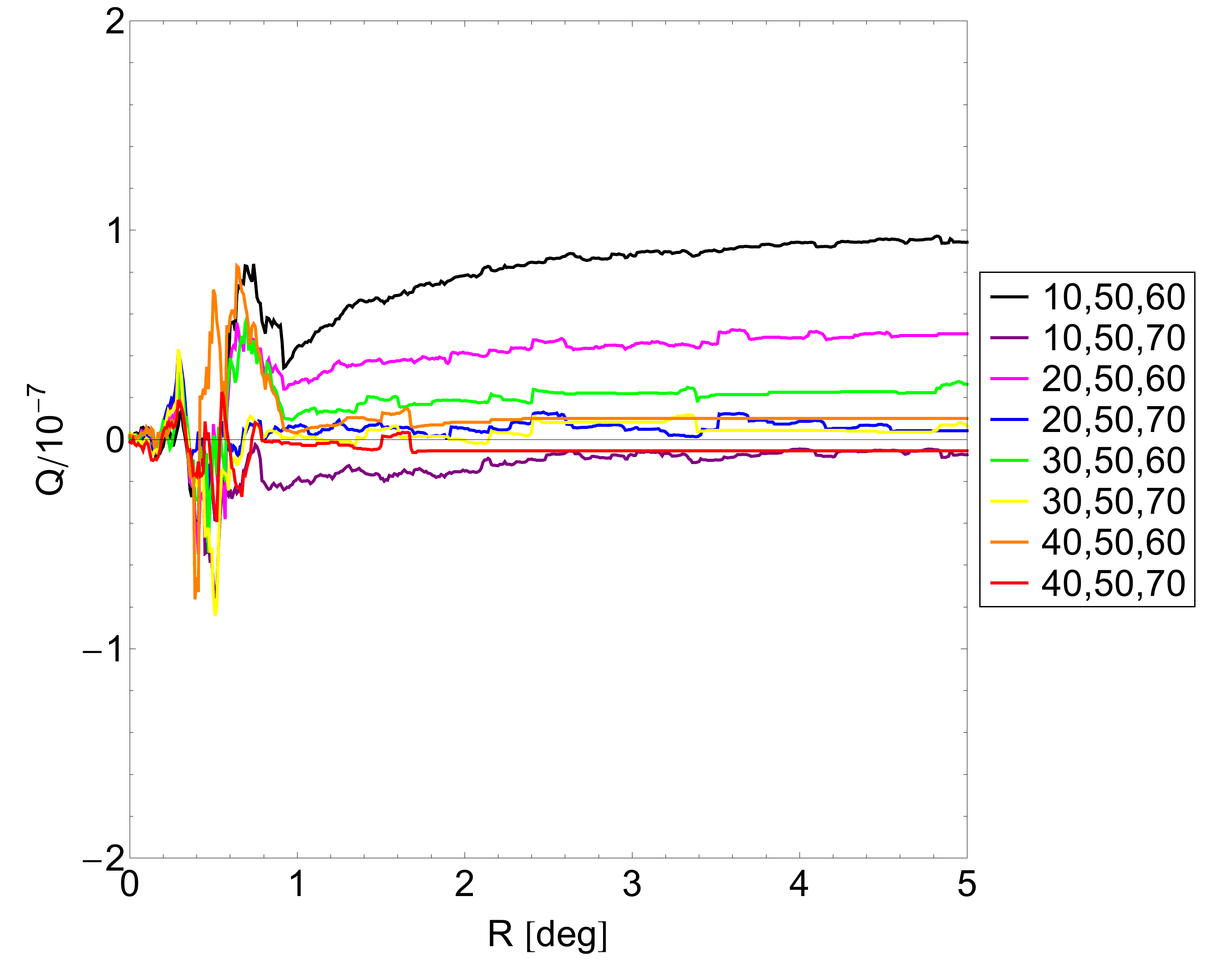}
\caption{$Q$-statistics for the case of a source tilted $5^\circ$ with respect to the line of sight, for $D_{\rm s} = 1\,{\rm Mpc}$, $E_{\rm TeV}=10\,{\rm TeV}$ and $\Psi = 5^\circ$.  All panels correspond to the three right hand panels of Fig.~\ref{fig:test5g}, {\it i.e.}~$f_{\rm H} = -1$ at the top, $f_{\rm H} = 0$ in the middle $f_{\rm H} = +1$ at the bottom panel. The triplets in the legends correspond to $E_{1},E_{2},E_{3}$ in GeV, each in intervals of $[E_{i},E_{i} + 10\,{\rm GeV}]$.
}
\label{fig:Qtest5g}
\end{figure*}

We now use Eq.~(\ref{Qstat}) to calculate the $Q$-factors for the three helicity scenarios analyzed ($f_{H} = -1$, 
$0$ and $1$). We display the results for the case of tilted jets ({\it i.e.}~the scenario shown in the right panel of 
Fig.~\ref{fig:test5g}) in Fig.~\ref{fig:Qtest5g}. We consider triplets of energies $(E_{1},E_{2},E_{3})$ as needed 
for Eq.~(\ref{Qstat}), where each energy $E_{i}$ corresponds to an interval $[E_{i},E_{i} + 10\,{\rm GeV}]$.

The reason we consider the scenario of {\em tilted} jets is that this is the most probable case -- it is
very unlikely for the blazar jet to be directed exactly along the line of sight.
As discussed in Ref.~\cite{Tashiro:2014gfa}, 
the function $Q(R)$ is expected to start at the origin since the angular deflections are
small for small $R$. For larger $R$, the magnetic helicity causes $Q$ to grow, and at much larger
$R$, $Q$ will approach a constant value ($Q_\infty$) as there are no more photons to include
at such large $R$. The large $R$ behavior gets modified in a realistic setting where, in addition to the
blazar photons, we also observe background photons from other sources. Then, for large $R$, the
blazar contribution gets diluted by the background noise and $Q$ decreases to zero. In this case,
we would see a peak in $Q(R)$ whose position is set by the relative number of blazar to background
photons. In our simulations, however, we do not include background photons and indeed find $Q \to Q_\infty$
at large $R$.

\subsection{Computing the $S$-Statistics}

As the last part of our results we present a new alternative way to quantify the pattern of gamma ray
arrival directions and thus, indirectly, the helicity orientation. The idea underlying idea of this new method that we denote $S$-Statistics (for ``Spiral'') is that a gamma ray from a cascade
that has a greater deflection away from the source direction due to the magnetic field will also have
a greater azimuthal deflection if the magnetic field is helical. The pattern of observed gamma rays
will have a spiral structure that can be measured by finding the average deflection of gamma rays, ${\bar \theta}(\phi$,$E_{\gamma}$),
as a function of the azimuthal angle $\phi$ and the considered gamma ray energy $E_{\gamma}$.
We assume that there is at least one angle $\phi_{\rm max}$ for which $\overline{\theta}(\phi,E_{\gamma})$ has a 
well-defined and significant maximum, {\it i.e.}~a maximum from an average in a bin 
which has a reasonable number of photons \emph{and} is statistically significant. 
We consider events inside a band around $\phi_{\rm max}$ with width $2\Delta\phi$,
{\it i.e.}~events with $\phi_{\rm max} - \Delta\phi \le \phi \le \phi_{\rm max} + \Delta\phi$. 
For a right-handed spiral there should be higher values of $\overline{\theta}(\phi,E_{\gamma})$ for $\phi < \phi_{\rm max}$ 
than for $\phi > \phi_{\rm max}$ inside the band, while for a left-handed pattern 
$\overline{\theta}(\phi,E_{\gamma})$ should be smaller for $\phi < \phi_{\rm max}$ than for $\phi > \phi_{\rm max}$.
In other words, the peak of the function ${\bar \theta}(\phi)$ should be skewed to the right or to the
left depending on whether the spiral is right- or left-handed, respectively.
By finding a measure for this asymmetry or skew of the maximum one can deduce the 
orientation and subsequently the magnetic helicity. 

More concretely, the calculation is performed in the following way: first, we subdivide the interval on 
which $\phi$ is defined, {\it i.e.}~$\phi \in [0,2\pi )$, 
in $n_{\rm bin}$ bins, such that each of the bins has a width $\delta\phi = 2\pi/n_{\rm bin}$.
The $j$th bin, which corresponds to the interval $[(j-1)\delta\phi, j\delta\phi)$, $j = 1,...,n_{\rm bin}$, will be 
labeled $\phi^{(j)} = (j-1)\delta\phi$. For each bin we calculate $\overline{\theta}$ by
\begin{equation} \label{thetabarphi}
\overline{\theta}(\phi^{(j)},E_{\gamma}) = \frac{1}{N_{j}} \sum_{\{i | \phi^{(j)} \le \phi_{i} < \phi^{(j + 1)} \}} \theta_{i}\,,
\end{equation}
where $(\phi_{i},\theta_{i})$ are the coordinates of the $i$th event in the set 
$\{i | \phi^{(j)} \le \phi_{i} < \phi^{(j + 1)} \}$ and $N_{j}$ is the total number 
of events in this set. If $\{i | \phi^{(j)} \le \phi_{i} < \phi^{(j + 1)} \}$ is empty, 
we set $\overline{\theta}(\phi^{(j)},E_{\gamma}) = 0$. Furthermore, for real data it might be necessary to restrict 
the analysis to events with $\theta$ 
smaller than a certain value $\theta_{\rm max}$ as for $\theta > \theta_{\rm max}$ background photons might 
dominate and result in a false signal.

In this set of $\overline{\theta}(\phi^{(j)},E_{\gamma})$ one now has to identify the relevant and significant maxima as well as the corresponding bin number $j_{\rm max}$ and calculate the 
quantities 
\begin{equation} \label{Phiplusminus}
\Phi_{-} = \sum_{j = j_{\rm max} - \delta_{\rm bin}}^{ j_{\rm max} - 1} \overline{\theta}(\phi^{(j)},E_{\gamma}) , ~
\Phi_{+} = \sum_{j = j_{\rm max} + 1}^{ j_{\rm max} + \delta_{\rm bin}} \overline{\theta}(\phi^{(j)},E_{\gamma}) ,
\end{equation}
where $\delta_{bin} \ge 1$ is the number of bins we need to consider in order to include the width of the peak.
Here one has to assume periodicity, {\it i.e.}~$\overline{\theta}(\phi^{(j + n_{\rm bin})}) = \overline{\theta}(\phi^{(j)},E_{\gamma})$.
Essentially $\Phi_{-}$ corresponds to the average value of ${\bar \theta}$ to the left of the peak and $\Phi_+$
to the right of the peak.

The final step is to define the $S$-statistics that measures the skewness of the peak
\begin{equation} \label{Sdef}
S \equiv \frac{\Phi_{-} - \Phi_{+}}{\Phi_{-} + \Phi_{+}}\,
\end{equation}
For a right-handed spiral $S$ will be positive, whereas for a 
left-handed spiral it will be negative.

\begin{figure*}
\includegraphics[scale=0.335]{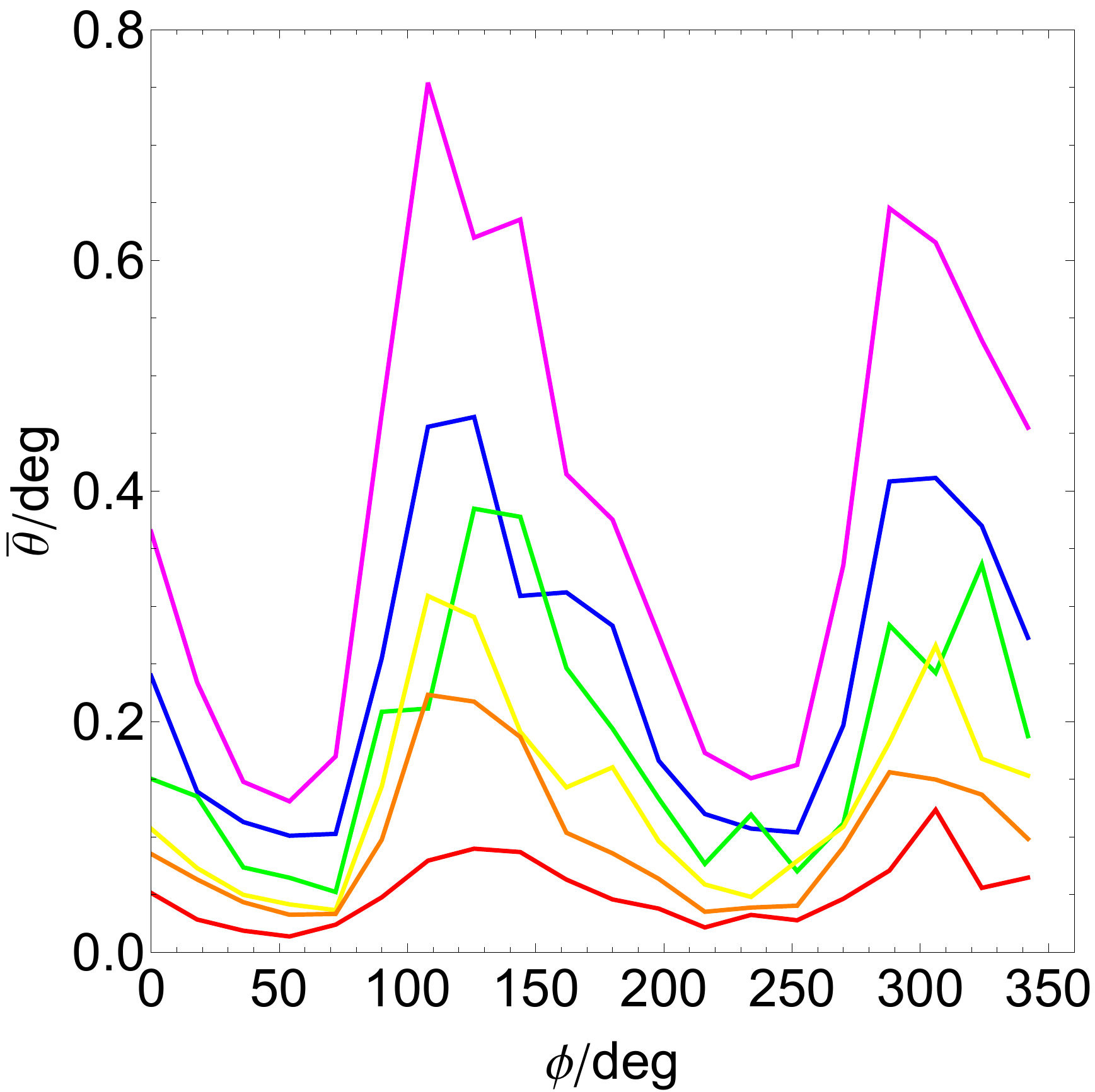}
\includegraphics[scale=0.335]{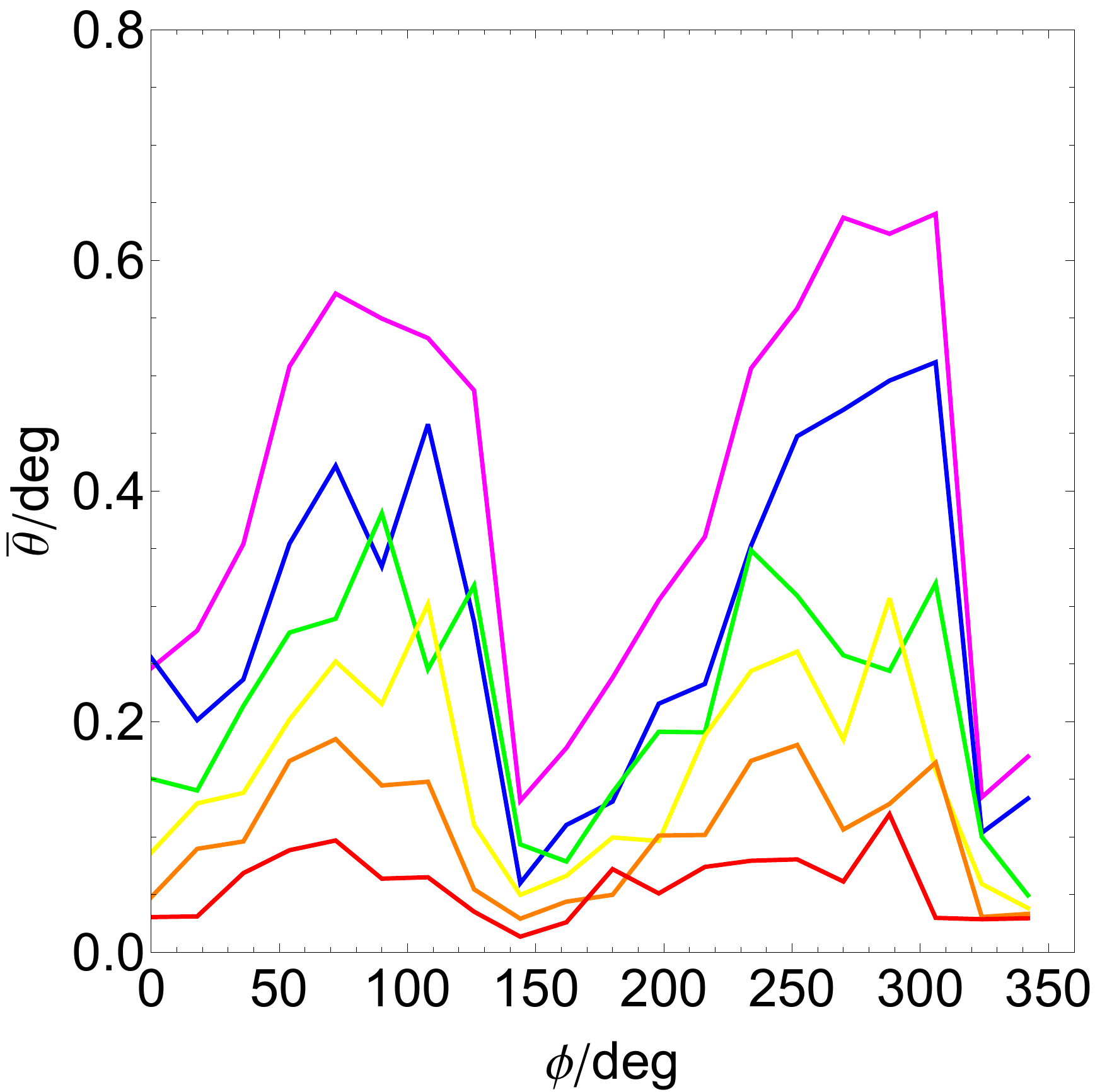}
\includegraphics[scale=0.335]{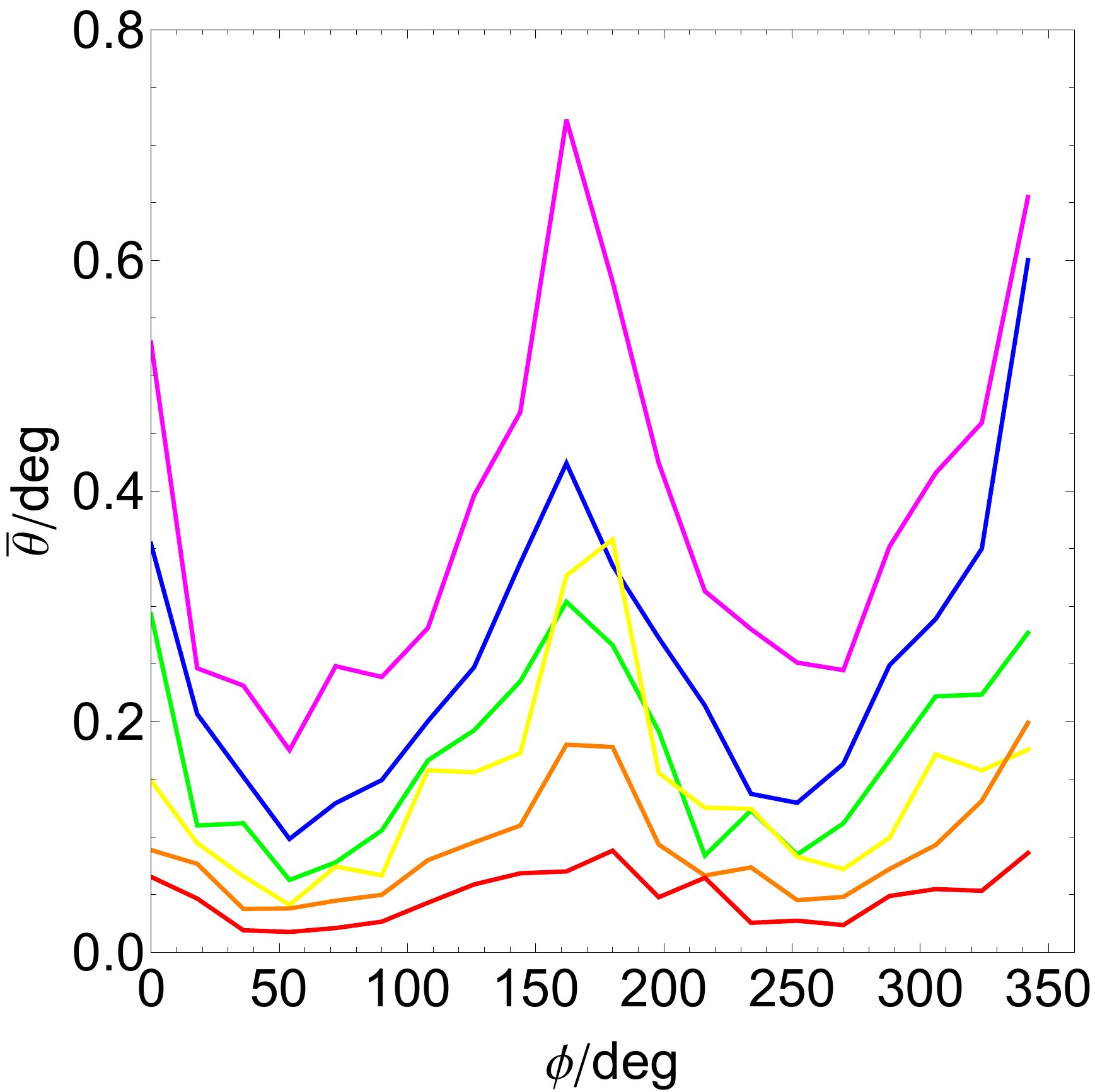}
\caption{The average polar angle $\overline{\theta}$ in dependence on the azimuthal angle $\phi$, as calculated in (\ref{thetabarphi}) for the three cases 
$f_{\rm H} = -1$ (left), $f_{\rm H} = 0$ (center) and $f_{\rm H} = +1$ (right), corresponding to the three cases in Fig.~\ref{fig:test5g}. Here we have chosen
$n_{\rm bin} = 20$. The colors correspond to the energy ranges of the arrival energies $E_{\gamma}$ in the same way as in Fig.~\ref{fig:HomField}, 
{\it i.e.}~$5 \-- 10\,{\rm GeV}$ (magenta), $10 \-- 15\,{\rm GeV}$ (blue), $15 \-- 20\,{\rm GeV}$ (green), $20 \-- 30\,{\rm GeV}$ (yellow), $30 \-- 50\,{\rm GeV}$ 
(orange), $50 \-- 100\,{\rm GeV}$ (red).}
\label{fig:thetabar}
\end{figure*}

We performed this computation for the data shown in Fig.~\ref{fig:test5g}. The plots for $\overline{\theta}(\phi,E_{\gamma})$, 
shown for different energies, are presented in Fig.~\ref{fig:thetabar}.
Even without any further analysis one can see in this figure that the peaks for opposite orientations indeed show 
opposite skews -- while on the
left panel higher angles are achieved for $\phi >\phi_{\rm max}$, on the right panel, even more clearly, that is the case for $\phi > \phi_{\rm max}$. For the 
central panel, however, peaks of either skewness are found.

In order to support these qualitative considerations, one has to look at the $S$-values which have been calculated and are presented for the three cases in 
Tabs.~\ref{tab:Stab-1}--\ref{tab:Stab+1}. The most clear case here is the one for $f_{\rm H} = -1$, where for all energy ranges we obtain $S < 0$ with two
values even going as low as $S \simeq -0.5$. This means that the morphology of the arrival directions is solely right-handed which is also clearly seen in the top 
panel of Fig.~\ref{fig:test5g}. For the case of $f_{\rm H} = +1$ the situation is less clear as there is only one value being as high as $S\simeq +0.48$. Nevertheless, 
since $S$ is positive for all energy ranges with three exceptions for which, however the absolute value of $S$ is close to zero, this is strong evidence for a left-handed 
orientation. Finally, no clear statement can be made regarding the case with no helicity ($f_{\rm H} = 0$) -- here one does not find significant negative or positive 
values for $S$.

As the last step, we need to connect the handedness of the arrival direction pattern with the sign of helicity. Their correlation has been found in Ref.~\cite{Long:2015bda}, where an analysis has been carried out for a homogeneous magnetic field. In this reference the authors indeed find that for a positive helicity one expects
a right-handed orientation, whereas for negative helicities a left-handed orientation should be observed, thus confirming our results for stochastic fields. 

As a concluding remark it should be stated that for the results presented above we used simulations containing approximately $1.4 \times 10^{5}$ 
photons arriving at Earth in the energy range $1.5 \le E_{\gamma}/{\rm GeV} \le 100$, which provided clear patterns with satisfactory statistical significance.
The upcoming Cherenkov Telescope Array (CTA)~\cite{2013APh....43....3A} might be able to detect this kind of signature in the energy range $E \simeq 10-100\;\text{GeV}$ with $\gtrsim 10\;\text{hours}$ of observation.
Fewer photons would distort the picture since, for example in the case of the $S$-statistics, the peaks would become less visible, such that a reliable 
calculation would no longer be possible. This is the case for high energy photons as their contribution to the total flux is rather small. On the other hand, 
for the lowest energies ($\sim\;$a few GeV), even with as few as $10^{4}$ photons relevant peaks can be seen, which, however, might be more difficult to 
construe in a more realistic case considering additionally diffuse gamma ray radiation and multiple sources.

\begin{table} 
\centering
\caption{Table for $S$ for maximal negative helicity ($f_{\rm H} = -1$).}
\begin{tabular}{cccccc}
\hline
\hline
 $E_{\gamma}$/GeV & $j_{\rm max}$ & $\phi_{\rm max}$/deg & $\Phi_{-}$ & $\Phi_{+}$ & $S$ \\
\hline
\hline
\multirow{2}{*}{5--10} & 6 & 108.0 & 0.768 & 1.67 & -0.37 \\ & 16 & 288.0 & 0.649 & 1.60 & -0.42  \\
\hline
\multirow{2}{*}{10--15} & 7 & 126.0 & 0.813 & 0.904 & -0.05 \\ & 17 & 306.0 & 0.709 & 0.882 & -0.11  \\
\hline
\multirow{2}{*}{15--20} & 7 & 126.0 & 0.472 & 0.818 & -0.27  \\ & 18 & 324.0 & 0.637 & 0.473 & 0.15  \\
\hline
\multirow{2}{*}{20--30} & 6 & 108.0 & 0.222 & 0.625 & -0.48  \\ & 17 & 306.0 & 0.370 & 0.428 & -0.07  \\
\hline
\multirow{2}{*}{30--50} & 6 & 108.0 & 0.163 & 0.507 & -0.51 \\ & 16 & 288.0 & 0.170 & 0.385 & -0.39  \\
\hline
\multirow{2}{*}{50--100} & 7 & 126.0 & 0.151 & 0.200 & -0.13  \\ & 17 & 306.0 & 0.145 & 0.172 & -0.09  \\
\hline
\hline
\end{tabular}
\label{tab:Stab-1}
\end{table}

\begin{table}
\centering
\caption{Table for $S$ for vanishing helicity ($f_{\rm H} = 0$).}
\begin{tabular}{cccccc}
\hline
\hline
 $E_{\gamma}$/GeV & $j_{\rm max}$ & $\phi_{\rm max}$/deg & $\Phi_{-}$ & $\Phi_{+}$ & $S$ \\
\hline
\hline
\multirow{2}{*}{5--10} & 4 & 72.0 & 1.69 & 12.12 & -0.11 \\ & 17 & 306.0 & 2.99 & 1.69 & +0.28  \\
\hline
\multirow{2}{*}{10--15} & 6 & 108.0 & 1.80 & 1.04 & +0.27 \\ & 17 & 306.0 & 2.21 & 1.28 & +0.27 \\
\hline
\multirow{2}{*}{15--20} & 5 & 90.0 & 1.12 & 1.07 & +0.02  \\ & 13 & 234.0 & 1.01 & 1.28 & -0.12  \\
\hline
\multirow{2}{*}{20--30} & 6 & 108.0 & 1.02 & 0.610 & +0.25  \\ & 16 & 288.0 & 1.07 & 0.608 & +0.28  \\
\hline
\multirow{2}{*}{30--50} & 4 & 72.0 & 0.463 & 0.470 & -0.01 \\ & 14 & 252.0 & 0.491 & 0.511 & -0.02  \\
\hline
\multirow{2}{*}{50--100} & 4 & 72.0 & 0.277 & 0.275 & +0.00  \\ & 16 & 288.0 & 0.418 & 0.218 & +0.31  \\
\hline
\hline
\end{tabular}
\label{tab:Stab0}
\end{table}

\begin{table} 
\centering
\caption{Table for $S$ for maximal positive helicity ($f_{\rm H} = +1$).}
\begin{tabular}{cccccc}
\hline
\hline
 $E_{\gamma}$/GeV & $j_{\rm max}$ & $\phi_{\rm max}$/deg & $\Phi_{-}$ & $\Phi_{+}$ & $S$ \\
\hline
\hline
\multirow{2}{*}{5--10} & 19 & 342.0 & 1.23 & 1.01 & +0.10 \\ & 9 & 162.0 & 1.15 & 1.32 & -0.07  \\
\hline
\multirow{2}{*}{10--15} & 19 & 342.0 & 0.888 & 0.713 & +0.11 \\ & 9 & 162.0 & 0.785 & 0.822 & -0.02  \\
\hline
\multirow{2}{*}{15--20} & 0 & 0.0 & 0.722 & 0.284 & +0.44  \\ & 9 & 162.0 & 0.594 & 0.542 & +0.05 \\
\hline
\multirow{2}{*}{20--30} & 19 & 342.0 & 0.428 & 0.309 & +0.16  \\ & 10 & 180.0 & 0.655 & 0.405 & +0.24  \\
\hline
\multirow{2}{*}{30--50} & 19 & 342.0 & 0.296 & 0.203 & +0.19 \\ & 9 & 162.0 & 0.285 & 0.338 & -0.08  \\
\hline
\multirow{2}{*}{50--100} & 19 & 342.0 & 0.157 & 0.131 & +0.09  \\ & 10 & 180.0 & 0.120 & 0.138 & +0.18  \\
\hline
\hline
\end{tabular}
\label{tab:Stab+1}
\end{table}

\section{Discussion and Outlook} \label{sec:Discussion}

We have performed three-dimensional Monte Carlo studies of the development of gamma-ray-induced electromagnetic cascades in the intergalactic medium in the presence of magnetic fields. 
We have used the ``Large Sphere Observer'' method for improved computational performance. In this case all cascade photons hitting the surface of the sphere are detected by the the observer. With a standard three-dimensional Monte Carlo simulation most cascade photons would not reach Earth, resulting in wasted computation and very low statistics.
A simplification made in our treatment is that the magnetic field evolves adiabatically with redshift as 
$B(z) = B(z=0) (1+z)^{2}$. This is justified because the cascade development we have discussed occurs 
in cosmic voids where MHD amplification and contamination by sources is minimal. Also, the sources
are at redshifts $z \lesssim 1$.
 
We first compared our computational setup
with analytical approximations and then validated it in simple scenarios containing a uniform magnetic field 
oriented parallelly and perpendicularly to the line of sight of the blazar jet. As expected, for a magnetic field parallel 
to the direction of the jet of half-opening angle $\Psi$, assumed to be pointing toward Earth, effects of the field 
were not observed. For a magnetic field perpendicular to the direction of the jet, deflections were non-zero and 
in the expected direction. Similar results were obtained for stronger and weaker magnetic fields and other 
orientations. These results are in accordance with Ref.~\cite{Long:2015bda} and also with the predictions 
of Eq.~(\ref{thetaEBD}). 

We have also studied the particular case of a magnetic field with a Batchelor power spectrum with and without helicity. The effects of helicity can be clearly 
seen in Fig.~\ref{fig:test5g}, where arrival directions follow right- or left-handed spirals, depending on the sign of the helicity. For stochastic fields, in 
general, the results tend to converge toward the case of a uniform magnetic field in the limit of large coherence lengths. We have considered only large values 
of correlation length ($L_{\rm c} \simeq 120\,{\rm Mpc}$) since for much smaller coherence lengths, with the other parameters being held fixed, no clear signature of helicity can be seen, as shown in Fig.~\ref{fig:diffLc}. 
Nevertheless, one should bear in mind that the current upper limits of coherence length of magnetic fields in 
voids range between a few and hundreds of Mpc~\cite{DuNe}, placing the chosen value of 120 Mpc well within the allowed bounds.

We have deployed the so-called $Q$-statistics, a powerful analysis tool that makes it possible to determine the properties of magnetic helicity directly from 
the observables of gamma rays measured at Earth. In this work we for the first time applied $Q$-statistics to realistic three-dimensional simulations of electromagnetic cascades. 
Our results for $Q$ are shown in Fig.~\ref{fig:Qtest5g}. The plots do not show a strong correlation
between $Q$ and the existence and sign of the helicity. At the moment we cannot clearly state whether averaging over
several objects will show a stronger correlation.
We plan on investigating this issue in a future work.

It is important to stress the fact that $Q$-statistics might not be the final method to quantify magnetic helicity, 
however it is a good initial approach and has been used in several works 
(Refs.~\cite{PhysRevD.87.123527,Tashiro:2013ita,Long:2015bda}) with satisfactory results.
In this work we have, for the first time, introduced the $S$-statistics, which is a direct measure of the handedness 
of a pattern with respect 
to the line of sight. We have shown that the orientation, represented by the sign of $S$, is directly correlated 
with the sign of helicity. 
This shows that the $S$ measure is also a powerful tool to be used in the analysis of helicity of IGMF.

Backgrounds at the $\sim\,$10-100 GeV energy range are expected due to secondary photons from AGN halos whose jet opening angles do not encompass the Earth. Other astrophysical sources of photons in this energy range also exist and have to be taken into account. In this first work we have neglected these backgrounds, which will be included in future studies.

We found that it is probably necessary to analyze various sources in order to make a definite statement about the sign of the helicity, since a clear signature cannot always be seen. In the future we will extend our simulations to the case of multiple sources and diffuse gamma rays. We expect to be able to reproduce actual detections and consequently retrieve more precise information about IGMF, which can be used to infer their origin and evolution.

In addition, we will extend the analysis by further exploring the parameter space as varying quantities such as the magnetic field strength 
$B_{\rm rms}$, the magnetic correlation length $L_{\rm c}$ and source parameters such as its distance from the observer, its energy spectrum or its cutoff energy, as they may 
be important in order to obtain a complete picture of their influence as discussed above and to explain actual observations.

\begin{acknowledgments}
R.~A.~B.~acknowledges the financial support from the John Templeton Foundation. The work of A.~S.~has been supported by the DAAD funded by the BMBF and the EU Marie 
Curie Actions. A.~S.~would like to thank the Arizona State University for the hospitality during his stay. Furthermore, A.~S.~is grateful to the "Helmholtz Alliance 
for Astroparticle Physics" (HAP) and the Collaborative Research Center SFB 676 "Particles, Strings and the Early Universe" for providing generous travel funds which were important
for the collaboration leading to this work. T.~V.~is supported by DOE, Office of High Energy 
Physics, under Award number \#DE-SC0013605 at ASU. T.~V.~is grateful to the Institute for Advanced Study, Princeton for 
hospitality while this work was being done. Special thanks go to Andrew J.~Long for his ideas and discussions which helped to complete this work and to 
NORDITA in Stockholm for organizing the workshop  ``Origin, Evolution, and Signatures of Cosmological Magnetic Fields'' during which important aspects of this 
work have been discussed.
\end{acknowledgments}

\end{document}